\documentclass{emulateapj}

\shorttitle{Dust poor quasars}
\shortauthors{Jun and Im}
\begin{document}
\title{Physical properties of luminous dust poor quasars}
\author{Hyunsung David Jun\altaffilmark{1} and Myungshin Im\altaffilmark{1}}
\altaffiltext{1}{Center for the Exploration of the Origin of the Universe (CEOU), Astronomy Program, 
Department of Physics and Astronomy, Seoul National University, Seoul 151-742, Republic of Korea;
hsjun@astro.snu.ac.kr, mim@astro.snu.ac.kr.}

\begin{abstract}
We identify and characterize a population of luminous dust poor quasars at 0$\,<$$\,z\,$$<\,$5, similar in photometric 
properties to the objects found at $z$$\,>\,$6 previously. This class of active galactic nuclei has been known to show 
little IR emission from a dusty structure, but is yet poorly understood in terms of number evolution or of dependence on 
physical quantities. In order to better understand the luminous dust poor quasar properties, we compiled a rest-frame UV 
to IR library of 41,000 optically selected type-1 quasars with $L_{\mathrm{bol}}$\,$>$\,$10^{45.7}\,$erg\,s$^{-1}$. After 
fitting the broad-band spectral energy distributions (SEDs) with accretion disk and dust components, we find 0.6\% of our 
sample to be hot dust poor with a rest-frame 2.3\,$\mu$m to 0.51\,$\mu$m flux density ratio of $-0.5$\,dex or less. 
The dust poor SEDs are blue in the UV--optical and weak in the MIR, such that their accretion disks are less obscured, and 
that hot dust emission traces that of warm dust down to the dust poor regime. At a given bolometric luminosity, dust poor 
quasars are lower in black hole mass and higher in Eddington ratio than general luminous quasars, suggesting that they 
are in a rapidly growing evolutionary state in which the dust poor phase appears as a short or rare phenomenon. The dust 
poor fraction increases with redshift, and possible implications for the evolution of the dust poor fraction are discussed. 
\end{abstract}

\section{Introduction}
Active galactic nuclei (hereafter AGNs) are known to be radiating through a wide range of wavelengths, where the unified 
model (e.g., \citealt{Kro88}; \citealt{Ant93}; \citealt{Urr95}) postulates dusty structures surrounding the central black hole 
to be responsible for the infrared emission. Thermally reprocessed emission coming from heated dust, is prominent in the 
infrared spectral energy distribution (hereafter SED) of AGNs. The hottest dust reaches temperatures up to 1500\,K and 
gives off near-infrared (hereafter NIR) excess radiation above the light from the accretion disk and the host galaxy, but 
the dust does not get much hotter due to sublimation of dust grains (e.g., \citealt{Bar87}). Hot dust in AGNs is common (e.g., 
\citealt{Gli06}), and is often modeled to lie in the inner dust torus (e.g., \citealt{Bar87}).

Recently, this common feature of NIR excess due to hot dust emission was found to be missing for a few quasars at 
$z$\,$\sim$\,6 (\citealt{Jia10}, hereafter J10). Based on the rest-frame NIR to optical flux ratios\footnote{We define 
$L_{\lambda}\,$=$\,L_{\lambda (\mu \mathrm{m})}\,$=$4\pi d_{L}^{2}\,\lambda F_{\lambda}$ as the rest-frame monochromatic 
luminosity, where $d_{L}$ is the luminosity distance and $F_{\lambda}$ is the rest-frame specific flux.}, $\log\,(L_{3.5}/L_{0.51})$, 
they found two quasars (out of 21 at $z$\,$\sim$\,6) with exceptionally small flux ratios such that the hot dust emission 
component is almost absent, while no such AGNs were found at lower redshifts. These particular examples were interpreted 
as the existence of quasars before the formation of dusty structure, observed in their early lifetimes. It was then by 
Hao et al. (2010, hereafter H10) when they selected hot dust poor quasars to show weak NIR continuum shapes from the 
diagram of NIR against optical slopes, and reported $\gtrsim\,$10\% of quasars to be hot dust poor even at low redshifts, 
although their definition of ``hot dust poor'' is different from that of J10. H10 provided wider possibilities in explaining 
the deficiency of hot dust radiation, raising dust destruction models as the origin. Meanwhile, Mor \& Trakhtenbrot (2011, 
hereafter M11) found 1.7\% of their quasars at redshifts of 0.75\,$<$\,$z$\,$<$\,2 to be hot dust free\footnote{Throughout 
this paper we use the term ``dust poor" to indicate weak IR dust emission below a certain threshold, and ``dust free'' to 
be completely absent of dust emission.}, having zero hot dust covering factors (hereafter CF$_{hd}$=$L_{hd}/L_{\mathrm{bol}}$), 
and $\gtrsim\,$16\% to be hot dust poor, lying below the lower $\sim$\,3\,$\sigma$ distribution of CF$_{hd}$. Still 
contrary to J10, they could not find a redshift dependence on the distribution of CF$_{hd}$, or on the fraction of their 
hot dust poor quasars. 

Not only the number evolution but also the origin of quasars to be hot dust poor, is yet controversial. Although J10 found hot 
dust poor quasars to have lower black hole masses ($M_{\mathrm{BH}}$) and higher Eddington ratios ($f_{\mathrm{Edd}}$), 
H10 could not locate their hot dust poor quasars to be clustered in a specific range of $M_{\mathrm{BH}}$, $f_{\mathrm{Edd}}$, 
or bolometric luminosity ($L_{\mathrm{bol}}$). Since both studies suffer from relatively small sample sizes of a few hundreds, 
a larger parent sample might help clarify the properties of dust poor quasars sizable in number. Based on 15,000 quasars 
with SDSS/WISE coverage, M11 still failed to find a meaningful link between the hot dust covering factor with $M_{\mathrm{BH}}$ 
or $f_{\mathrm{Edd}}$, and from the scatter in the correlations suggested that CF$_{hd}$ is independent on the evolutionary 
stage of the black hole. The latest study of \citet{Ma13} about the properties of warm dust covering factors (CF$_{wd}$) 
of 12,000 quasars, agrees with M11 for the anti-correlation between CF$_{wd}$ and $L_{\mathrm{bol}}$, but disagrees for the 
anti-correlation in CF$_{wd}$ and $M_{\mathrm{BH}}$, further complicating the case.

Previous studies on dust poor quasars follow mutually different definitions of being hot dust poor/free, thus adding an 
uncertainty when trying to compare the results on each of selected dust poor populations. For luminous quasars, the 
selection criterion of J10 based on the NIR-to-optical flux ratio, verifies 
that it picks out SEDs with an apparently weak NIR bump originating from the hot dust radiation. On the other hand, for 
less luminous ($L_{\mathrm{bol}}\lesssim10^{46}\,$erg\,s$^{-1}$) quasars where host galaxy contamination to the SED 
becomes meaningful, it would be better to use dust emission strength indicators taking into account the fraction of host 
galaxy light (H10), or to apply a luminosity dependent average galaxy contamination correction to the observed fluxes 
(e.g., \citealt{She11}, hereafter S11). Moreover, the selection of hot dust poor quasars up to date relies on the weak NIR 
emission, and it is yet unclear whether hot dust poor quasars are also dust poor in warmer phases, where the mid-infrared 
(hereafter MIR) is thought to be the peak wavelength of AGN dust emission (e.g., \citealt{Ric06}). Lastly, all previous 
studies lack either the sample size (J10, H10) or redshift coverage (M11, \citealt{Ma13}) to meaningfully disentangle 
number evolution and physical parameter characteristics of dust poor quasars.

In this paper, we aim to identify the lower redshift counterpart to the high redshift hot dust poor quasars of J10, to compare 
and understand the observational features of local to high redshift populations. Where our sample quasars are defined to 
be luminous enough to have almost negligible host contamination (\S2), we choose NIR-to-optical flux ratios to find weak 
hot dust emission sources matching with those in J10. Our spectro-photometric data (\S2) of 41,000 quasars at 
0$\,<$$\,z\,$$<\,$5 with contiguous wavelength coverage from UV to IR, enables reliable modeling of the SED and 
measurement of both hot and warm dust emission strengths (\S3). With the help of large number statistics and multi-wavelength 
data, we are able to better identify dust poor quasars, constrain the evolution in the redshift space, and understand 
their nature reflected through key physical parameters (\S4, \S5). Throughout this paper we adopt a flat $\Lambda$CDM 
cosmology with parameters of $H_{0}=\mathrm{70\,km\,s^{-1}\,Mpc^{-1}}$, $\Omega_{m}=0.3$, and $\Omega_{\Lambda}=0.7$.

\section{Sample definition and AGN data set}
For the selection of hot dust poor quasars we follow J10 to use NIR-to-optical flux ratios, in order to indicate the relative 
strength of the dust emission. Defining $f_{\lambda}$ as the $\lambda\,\mu$m-to-optical flux ratio\footnote{We use rest-frame 
flux for $F_{\lambda}$, $F_{\nu}$ throughout this paper, in upper cases to prevent confusion with the rest-frame flux 
ratio $f_{\lambda}$.}, \begin{equation} f_{\lambda}\,=\,\log\,(\lambda F_{\lambda}/{0.51 F_{0.51}})\,=\,\log\,(L_{\lambda}/L_{0.51}), \end{equation} 
we assign ``hot dust poor'' quasars to satisfy $f_{2.3}$\,$<$\,$-0.5$. $f_{2.3}$ is chosen for the 2.3\,$\mu$m data to 
be effective in constraining the hot dust emission, as it is the wavelength where the 1250\,K black body component for the 
SED fitting (\S3), peaks in the $F_{\lambda}$ space. In addition, the hot dust poor criterion is set to select objects with 
a weak NIR bump at 2.3\,$\mu$m, equivalent to J10 objects that are below the lower 3\,$\sigma$ distribution in $f_{3.5}$ 
($\lesssim$\,$-0.5$). In Figure 1, our selection criterion is compared to SEDs of quasars in J10. Due to the variety of quasar 
optical continuum slopes, a single value of $f_{2.3}$ can be derived from a range of $\alpha$, in $F_{\nu} \propto \nu^{\alpha}$. 
In the figure, we show two example SEDs of different continuum slopes that satisfy $f_{2.3}$\,=\,$-0.5$, with $\alpha$ 
lying within 1\,$\sigma$ to the average of all quasars in this work. If $\alpha$\,=\,$-0.24$, the SED (black solid line) is 
purely power-law in the optical--NIR, and when $\alpha$\,=\,0.10 the SED (black dashed line) is the sum of a power-law, 
plus a weak hot dust emission component similar in strength to that of J1411+1217 from J10. The $f_{3.5}$ for both examples 
are $-0.64$ and $-0.57$, which are higher than $f_{3.5}$\,$\lesssim$\,$-1$ for the two hot dust free quasars in J10, but 
still below the lower 3\,$\sigma$ range of $f_{3.5}$ in Figure 2 of J10, separating weak hot dust radiating AGNs from the 
rest of the distribution. 

\begin{figure}
\centering
\includegraphics[scale=.9]{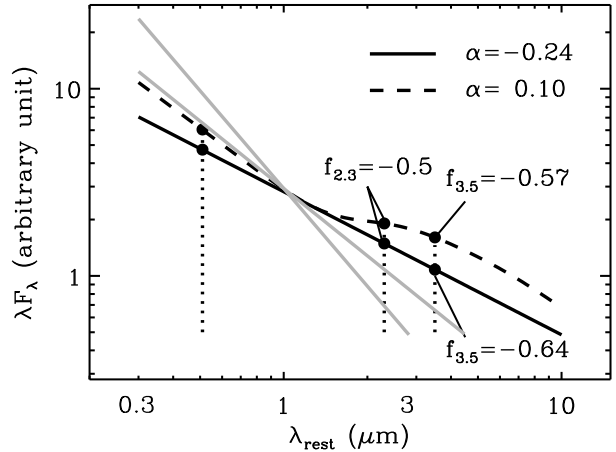}
\caption{Graphical examples of two hot dust poor quasar SEDs (black lines) with $f_{2.3}$\,=\,$-0.5$ from Equation 1, where 
$f_{2.3}$ are calculated from the logarithmic ratio of 2.3\,$\mu$m to 0.51\,$\mu$m fluxes in $\lambda f_{\lambda}$ 
space. These cases falling on the marginal limit of hot dust poor quasar selection, correspond to SEDs that are either purely 
power-law (black solid line) or power-law plus weak hot dust emission (black dashed line), with both $f_{3.5}$ values lying 
below 3\,$\sigma$ of the distribution in J10. Overplotted are the two extreme SEDs comparable in $f_{3.5}$ to dust free 
quasars in J10 (gray lines). \label{fig1}} 
\end{figure}

\begin{figure*}
\centering
\includegraphics[scale=.9]{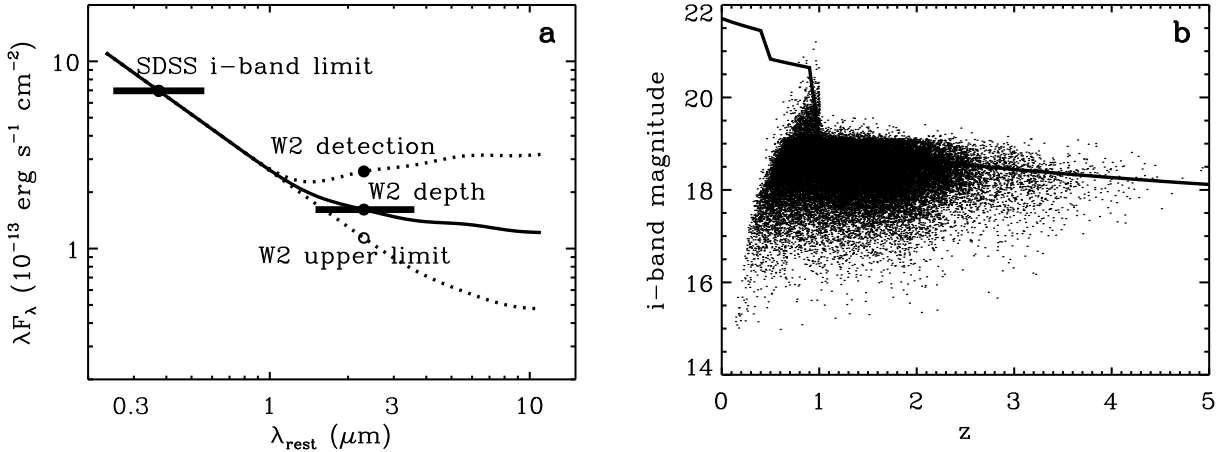}
\caption{Left: An illustration of Sloan $i$-band limit for a marginal hot dust poor quasar with $f_{2.3}$\,=\,$-0.5$ 
at $z$\,=\,1, set for the rest-frame 2.3\,$\mu$m to fall on the WISE (in this case $W$2) S/N=2 detection limit. Under the 
$i$-band flux limit not only hot dust rich ($f_{2.3}>-0.5$) objects are detected at 2.3\,$\mu$m, hot dust poor 
($f_{2.3}$\,$<$\,$-0.5$) objects are either detected or given with upper limits deeper than the 2.3\,$\mu$m flux that meets 
$f_{2.3}$\,=\,$-0.5$, such that the classification of hot dust poor$/$rich quasars becomes more complete under the given 
WISE sensitivity. 
Right: The SDSS $i$-magnitude cut (solid line) as a function of redshift. $f_{2.3}$\,=\,$-0.5$ quasars on this line, will 
fall on the WISE rest-frame 2.3\,$\mu$m detection limit for $t_{exp}$\,=\,97\,s. Overplotted dots are objects satisfying 
the $i$-band sensitivity limit for their respective WISE limits, mostly preserving the main magnitude limit ($i$\,$<$\,19.1) 
of SDSS quasar selection. Objects above the solid line remain in the sample since their WISE image depths are deeper than 
the fiducial WISE limit. \label{fig2}} 
\end{figure*}

The search for hot dust poor quasars is based on the Sloan Digital Sky Survey quasar catalog (DR9 version, \citealt{Sch10}; 
\citealt{Par12}), which provides the largest number (185,801) of optically selected type-1 AGNs together with ancillary 
AGN properties (S11), from a survey area of 14,555\,deg$^{2}$. In addition to the SDSS optical imaging and spectroscopic 
data, we gathered UV through MIR imaging data from GALEX, 2MASS, UKIDSS and WISE surveys (\citealt{Mar05}; \citealt{Skr06}; 
\citealt{Law07}; \citealt{Wri10}) covering $\sim$\,26,000\,deg$^{2}$, all-sky, $\sim$\,4,000\,deg$^{2}$, and all-sky, 
respectively. The multi-wavelength data set enables the measurement of hot and warm dust emission strengths, $f_{2.3}$ 
and $f_{10}$, where $f_{10}$ is to be used to probe the warm dust emission of hot dust poor quasars in comparison with 
$f_{2.3}$. Also, additional NIR data were obtained from our own UKIRT imaging observations, to improve the photometric 
accuracy of 113 type-1 AGNs that have low S/N in the 2MASS data, including 32 hot dust poor objects at $z\gtrsim2$. Out 
of the hot dust poor quasars observed by UKIRT, 19 satisfy the criterion of $f_{2.3}$\,$<$\,$-0.5$ while 13 objects do not.

Objects in the SDSS quasar catalog were matched within 2\,$\arcsec$ radii, with multi-wavelength data from UV through MIR: 
GALEX GR7 for UV, 2MASS and UKIDSS DR10 for NIR, and WISE all-sky for MIR. Since the GALEX and UKIDSS data consist of 
several separate surveys (AIS, MIS, DIS, NGS, GII, CAI for GALEX, and LAS, GCS, DXS for UKIDSS), we merged each catalog
consisting the UV and NIR survey data respectively, with overlapping sources to have a single photometry of better quality. 
The matching was performed for all the UV--MIR data set, but for the 2MASS data, we instead took the matched catalog in 
\citet{Sch10}, which adopts the closest neighbor for multiple matches of 1.5\% occurrence. On the other hand, multiple 
SDSS to UKIDSS NIR matches of 0.72\% occurrence were removed instead of keeping the closest pair, to prevent source 
confusion in UV/MIR data under poor seeing, where UKIDSS data have the best seeing to tell whether or not there is a 
confusion. In the absence of UKIDSS coverage the number of multiple SDSS-2MASS matches could not be fully trusted, due to 
the higher incidence of multiple matching than that of SDSS-UKIDSS at an identical 2\,$\arcsec$ matching radius. For this 
reason, we did not reject the multiple SDSS-2MASS matched objects without UKIDSS coverage in principle, but visually 
re-examined the SDSS images of the vicinity of hot dust poor quasars (\S4.1), to remove those with definite source confusion 
within the angular resolution of SDSS imaging. Meanwhile, hot dust rich ($f_{2.3}$\,$>$\,$-0.5$) objects without UKIDSS 
data are exposed under a mild level of confusion, expected to be 0.72\% from the SDSS-UKIDSS matching, though we regard 
this fraction to be negligible in counting the number of hot dust rich quasars later on. Finally, objects undetected or 
blended in WISE were given with upper flux limits, while those undetected in the NIR were removed, in order to provide 
careful constraints on the dust emission strength.

The photometric data collected from multi-wavelength catalogs can potentially suffer from the following problems. First, 
the photometry of AGNs include contributions from the host galaxy that needs to be minimized or subtracted somehow. 
Second, the mixture of different definitions of Kron, PSF, aperture, and profile fit magnitudes for UV, optical, NIR, and 
MIR data, brings inaccurate photometry of extended objects from PSF or aperture magnitudes. Both of these potential 
problems can be avoided by minimizing the host galaxy contamination. Thus, we limited our sample with a bolometric 
luminosity cut of $L_{\mathrm{bol}}>10^{45.70}\,$erg\,s$^{-1}$, derived from $L_{0.51}$\,$>$\,$10^{44.73}\,$erg\,s$^{-1}$ 
(\S3) using the optical to bolometric correction of 9.26 (S11). By doing so, host contamination is limited to less than 
10\% in 5100\,\AA (S11). After applying the luminosity cut, we double checked if fixed-aperture (PSF or aperture) 
magnitudes bring consistent photometry to total magnitudes by comparing PSF versus Petrosian magnitudes for SDSS $ugriz$ 
filters, and aperture versus Petrosian magnitudes for 2MASS or UKIDSS $JHK$ filters. We found the median difference 
between the magnitude systems to be at most 0.03 mag at all filters, with the rms scatter to fall within 0.1 mag for 
SDSS or UKIDSS data, and within 0.2 mag for that of 2MASS. Therefore we consider our luminosity cut sample to have a 
compatible set of magnitudes dominated by the central AGN contribution.

As a next step, we imposed an $i$-band flux limit that varies with redshift and the WISE depth for each object. This is 
necessary since the WISE flux limits are not deep enough for some of the SDSS quasars to determine if the object is 
$f_{2.3}$\,$<$\,$-0.5$. In other words, we selected objects that are bright enough in both observed-frame $i$- and WISE 
bands, to allow us to determine if the object is dust poor or not. Since each observed band traces different rest-frame 
wavelength at different redshift, the $i$-band cut changes as a function of redshift. Furthermore, the WISE depths are 
not uniform over the entire sky, so that the $i$-band depth needs to be different at each location on the sky. The $i$-band 
flux limits were determined by imposing a quasar with $f_{2.3}$\,=\,$-0.5$ at the $i$-magnitude, to have 2\,$\sigma$ 
detection in the WISE band that covers the rest-frame 2.3\,$\mu$m. Figure 2a shows a fiducial $i$-band magnitude limit 
which assumes the typical WISE limit with 97\,s exposure (11 frames, \citealt{Wri10}). When the exposure time, 
$t_{exp}$, was different at the position of another quasar, we adjusted the fiducial $i$-band limit by adding 
1.25\,$\log$($t_{exp}$/97\,s). The objects passing the $i$-band limit are depicted in Figure 2b. 

Lastly, we required at least two data points to lie in the rest-frame 0.3--1\,$\mu$m with one of the points at shorter 
than 0.6\,$\mu$m, so that the optical continuum slope could be reliably measured from the SED fitting. In short, bookkeeping 
the number of matches through the sample selection process, 100\%, 54\% of SDSS quasars have NIR coverage and detection, 
where almost all the undetections come from the shallow 2MASS data. From then, 100\%, 91\% are covered and detected 
in the WISE MIR, yielding 49\% of the initial catalog to be matched with contiguous rest-frame UV--NIR information, of 
which 82\% are matched in the GALEX UV. The bolometric luminosity cut passes through 81\% of the multi-wavelength data, 
while the $i$-band limit leaves a further 56\% of the remaining sample. Finally, the constraints on the shallow WISE upper 
limits and the number of optical data points sum up to a 2\% rejection, leaving 22\% of the initial SDSS sample or 40,825 
objects, for the SED fitting analysis. We summarize the final data set and observations in Table 1.

\section{Fitting of broad-band SED and spectra}
We modeled individual quasar SEDs within rest-frame 0.3--20\,$\mu$m boundaries, as a combination of power-law continuum 
feature from the accretion disk, and black body emission from the heated dust in hot and warm phases. For all quasars a default 
1250\,K hot dust emission was considered because it well describes the NIR part of the SED \citep{Gli06}. In cases where the 
rest-frame 3.5\,$\mu$m and 9\,$\mu$m were available, 500\,K and 200\,K black bodies were added respectively to model the 
MIR continuum emission of AGNs through warm dust phases, where the combination of warm dust temperatures are found to 
well fit the observed SEDs of AGNs (e.g., \citealt{Bar87}; \citealt{Hao05}). Quantitatively expressing the model SED as 
\begin{equation} F_{\lambda}=F_{\lambda,disk}+F_{\lambda,dust}, \end{equation}
it consists of an accretion disk component, $F_{\lambda,disk}$\,=\,$c_{disk}\,\lambda^{-(2+\alpha)}$, and a combination 
of dust emission components, $F_{\lambda,dust}$\,=\,$c_{hd}\,B_{\lambda}(1250\,\mathrm{K})$ + $c_{id}\,B_{\lambda}(500\,\mathrm{K})$ 
+ $c_{wd}\,B_{\lambda}(200\,\mathrm{K})$, where $\alpha$ is the power-law continuum slope in $F_{\nu} \propto \nu^{\alpha}$, 
and $c_{hd}, c_{id}, c_{wd}$ are contributions from the hot (1250\,K), intermediate (500\,K), and warm (200\,K) dust black 
bodies. $c_{id}, c_{wd}$ were used when the rest-frame IR data included 3.5\,$\mu$m/9\,$\mu$m, the geometric mean of the 
peak wavelengths out of hot--intermediate/intermediate--warm components, while the coefficients were fixed to zero otherwise. 
The peak of the black body radiation from hot and warm dust models in $F_{\lambda}$ space are at 2.3 and 14\,$\mu$m each, 
which means that WISE covers hot dust emission mostly within $W$1--$W$3 bands over $z$\,=\,0--4, while warm dust emission 
are probed usually under the inclusion of $W$4 at $z\lesssim$\,0.5. 

\begin{deluxetable}{cccc}
\tablecolumns{4}
%\tabletypesize{\scriptsize}
\tablecaption{Summary of imaging data}
\tablewidth{0.47\textwidth}
\tablehead{
\colhead{Name} & \colhead{Filters} & \colhead{$N$} & \colhead{$t_{exp}$(s)}}
\startdata
\sidehead{Cataloged data}
GALEX	  & FUV/NUV & 17,797/33,741 & 368/1,504\\
SDSS      & $ugriz$   & 40,825 & 54\\
2MASS   & $JHK$     & 33,725 & 8\\ 
UKIDSS  & $YJHK$    & 18,698 & 40\\
WISE      & $W1$--$W4$ & 40,825 & 123\\
\sidehead{Newly acquired data}
UKIRT     & $YJHK$   & 113   & 100
\enddata
\tablecomments{N is the number of objects detected in at least one filter within the corresponding data set, out of the 
N=40,825 final multi-wavelength matched sample limited in bolometric luminosity, $i$-band flux, and the number of optical 
data points, from the initial N=185,801 SDSS quasars. $t_{exp}$ is the median exposure time of a single-band image in 
the corresponding data set.} 
\end{deluxetable} 

The SED fitting was performed over the entire sample of 40,825 objects, which excludes 139 and 895 quasars that were 
rejected (\S2) due to shallow WISE upper limits or insufficient rest-frame optical coverage, respectively. The median 
reduced chi-square $\chi^{2}_{\nu}$, are not the best at 5.4 and 3.0, for all and hot dust poor objects separately. 
Still, we find the large $\chi^{2}_{\nu}$ sources (\S4.1) to show acceptable fits to the data, while there are no cases 
without a fitting solution within the entire sample. A rather large $\chi^{2}_{\nu}$ values do not indicate the poor 
determination of the continuum slope, since they are caused by wiggly features in the continuum such as broad emission 
lines or FeII complex, and photometric variabilities between different wavelength data sets. A further test on the accuracy 
of the SED fitting under special conditions, when the gap between the WISE $W$2 and $W$3 bands becomes problematic in 
constraining the $f_{2.3}$, is described in \S4.2.

Having gone through the SED fitting, we used IR-to-optical flux ratios $f_{2.3}, f_{10}$, to quantify hot/warm dust emission 
strengths over that of the optical continuum. We measured $f_{2.3}, f_{10}$ from the fluxes on the fitted SED curve at the 
corresponding rest-frame wavelengths, while $f_{10}$ was computed only when the warm (200\,K) dust component was used 
for the SED fit. Upper limits to $f_{2.3}, f_{10}$ were provided from the upper flux limits for the case of WISE undetection, 
assuming the upper flux limits as detections. The use of $f_{2.3}, f_{10}$ instead of $\alpha, c_{disk}, c_{hd}, c_{id}, c_{wd}$, 
is more straightforward to select quasars with weak dust emission, as $c_{hd}, c_{id}, c_{wd}$ are easily coupled with $\alpha$ 
which varies by object. An example is a red quasar small in $\alpha$ due to optical extinction, and large in $f_{2.3}, f_{10}$ 
from strong hot--warm dust emission. Because the extrapolated red power-law component could take care of the NIR--MIR 
emission, the resultant small $c_{hd}, c_{id}, c_{wd}$ for instance, are not necessarily good indicators of little dust emission. 

We wanted to pay careful attention for the modeling to be robust under systematic effects. First, broad emission lines 
could disturb the optical broad-band fluxes from a simple power-law continuum model, where H$\alpha$ is the strongest 
and the only meaningful ($>$\,0.02 dex) line contamination within our wavelength of interest \citep{Hao11}. Therefore we removed the 
broad-band point enclosing rest-frame H$\alpha$, if the $\chi^{2}_{\nu}$ containing that data point became larger than 
that without. We additionally note that, although the rest-frame 10\,$\mu$m is often surrounded by polycyclic aromatic 
hydrocarbon emission and silicate absorption features, we expect the line contaminations to be negligible considering the 
very wide filter transmission of $W$3 and $W$4. Second, variability can be a problem with multi-wavelength data taken 
at different epochs, for it would possibly bring flux offsets. As a sanity check we computed the scatter in the magnitude 
difference between the 2MASS versus UKIDSS $J$ and $K$ AB magnitudes, to find the median rms of the magnitude difference 
to be $\sim$\,0.1 mag after subtracting the magnitude measurement error in quadrature. Because this level of variability 
is limited, and indistinguishable with that derived from hot dust poor quasars only, we consider the variability issue to 
be tolerable for the selection of hot dust poor SEDs. 

To supplement the photometric products with that from spectroscopy, we compiled black hole masses based on the UV/optical 
line and continuum fitting by S11 (of both DR7 and DR9 data), where we adopted $M_{\mathrm{BH}}$ estimators in the form of
\begin{equation}
\log \Big(\frac{M_{\mathrm{BH}}}{M_{\odot}}\Big)=a+b\log\Big(\frac{L_{\lambda}}{10^{44} \,\mathrm{erg}\,\mathrm{s}^{-1}}\Big)+c\log\Big(\frac{FWHM}{10^{3}\,\mathrm{km}\,\mathrm{s}^{-1}}\Big), \end{equation} 
with $L_{\lambda}$\,=\,$(L_{0.51}, L_{0.51}, L_{0.135})$ and $(a, b, c)$=\,\{(6.69, 0.50, 2.1),\,(6.91, 0.50, 2),\,(6.66, 0.53, 2)\}, 
for spectral regions around (H$\alpha$, H$\beta$, CIV) respectively. For the MgII estimator $L_{\lambda}$\,=\,$L_{0.3}$ and 
$(b, c)$\,=\,(0.62, 2), while $a$\,=(6.75, 6.81, 6.79) depending on the narrow line to be subtracted/included (DR7), or 
unused for the fitting (DR9), individually. Each (H$\alpha$, H$\beta$, MgII, CIV) recipe was originally from \citet{Gre05}, 
\citet{Ves06}, \citet{Mcl04}, and \citet{Ves06}, but we primarily followed the cross calibrated form of $(a, b)$ from S11 to 
provide consistency between $M_{\mathrm{BH}}$ from different lines. Within the following redshift intervals, we used the black 
hole mass estimator of H$\alpha \,(z$\,$<$\,$0.37)$, H$\beta \,(z$\,$<$\,$0.84)$, MgII$\,(0.7$\,$<$\,$z$\,$<$\,2.1), 
and CIV$\,(z$\,$>$\,2.1), while masses from different estimators were averaged in overlapping redshifts.

\begin{figure}
\centering
\includegraphics[scale=.9]{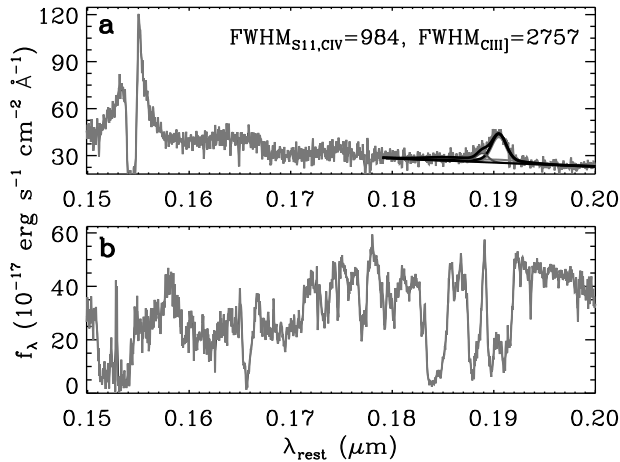}
\caption{Top: An example of CIV$\lambda$1549 broad line suffering from an absorption feature, and the CIII]$\lambda$1908 
spectral region with emission line fits around the CIII] overplotted. In this case where the CIII] is well identified from 
neighbor lines, the CIV FWHM is replaced with that of CIII]. The broad line FWHM (in km s$^{-1}$) from S11 and from our 
fits, are printed. Bottom: An example of CIV and CIII] both showing BAL features, where the $M_{\mathrm{BH}}$ information 
is dropped. \label{fig3}}
\end{figure}
\begin{figure}
\centering
\includegraphics[scale=0.9]{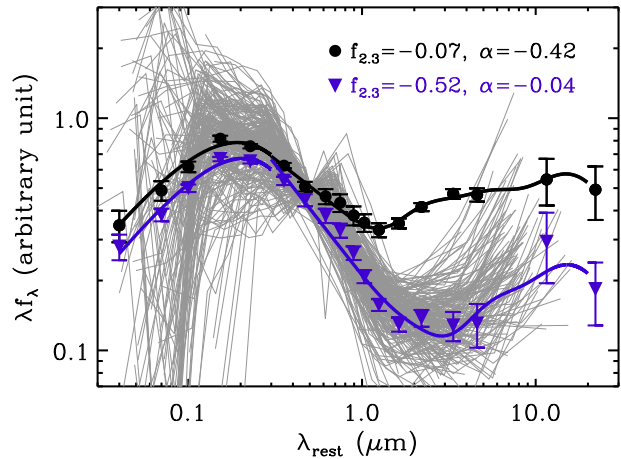}
\caption{Composite SED of hot dust poor (purple, downward triangles) and all (black circles) quasars, gridded on the rest-frame 
400, 700, 1000\,\AA--GALEX--SDSS--UKIDSS--WISE wavelengths. Solid lines are polynomial fits for $<$\,0.3\,$\mu$m, and 
power-law continuum+black body dust fits at $>$\,0.3\,$\mu$m. Individual hot dust poor SEDs based on the detected data 
points are overplotted with thin gray lines, while we note that the MIR part are generally poorer in sensitivity due to shallow 
$W$3 and $W$4 imaging. \label{fig4}}
\end{figure}

Due to the automatic nature of the spectral fitting in S11, we visually inspected the fits for hot dust poor quasars as a 
sanity check, tried direct fitting by ourselves, and compared the $M_{\mathrm{BH}}$ in S11 from different methodologies. 
The continuum and line fits by S11 were reproducible overall, but with cases of problematic fitted results. First of all, we 
checked the H$\beta$ fits by following the method of S11 except for using the FeII template of \citet{Tsu06}, to fit the 
spectra around the H$\beta$ region. For a total of 9 quasars at $z$\,$<$\,$0.84$, we found a good agreement in our H$\beta$ 
$M_{\mathrm{BH}}$ to S11 values, to have a $-0.09$\,$\pm$\,0.08\,dex relative offset and scatter. Thus, we trust and keep 
the H$\beta$ $M_{\mathrm{BH}}$ of S11. Likewise, we performed CIV fits of 92 hot dust poor quasars at $z>$\,2.1 with 
similar methodology to S11, but carefully masking out the absorption features. We found 20 spectra with strong absorption 
including broad absorption line (hereafter BAL) systems, such that the fitting would be meaningless. Hence, for the spectra 
with severe CIV absorption but good quality in CIII]$\lambda$1908 (e.g., Figure 3), we fitted the CIII] to use the line width 
as an effective CIV FWHM surrogate \citep{She12} for 9 objects, while we dropped the $M_{\mathrm{BH}}$ of remaining 11 
objects. The CIV $M_{\mathrm{BH}}$ without absorption spectra, are offset by $-0.05$\,$\pm$\,0.22\,dex to S11 values, 
where we keep our CIV and CIII] based $M_{\mathrm{BH}}$ for the better treatment of absorption features.

\begin{figure*}
\centering
\includegraphics[scale=0.5]{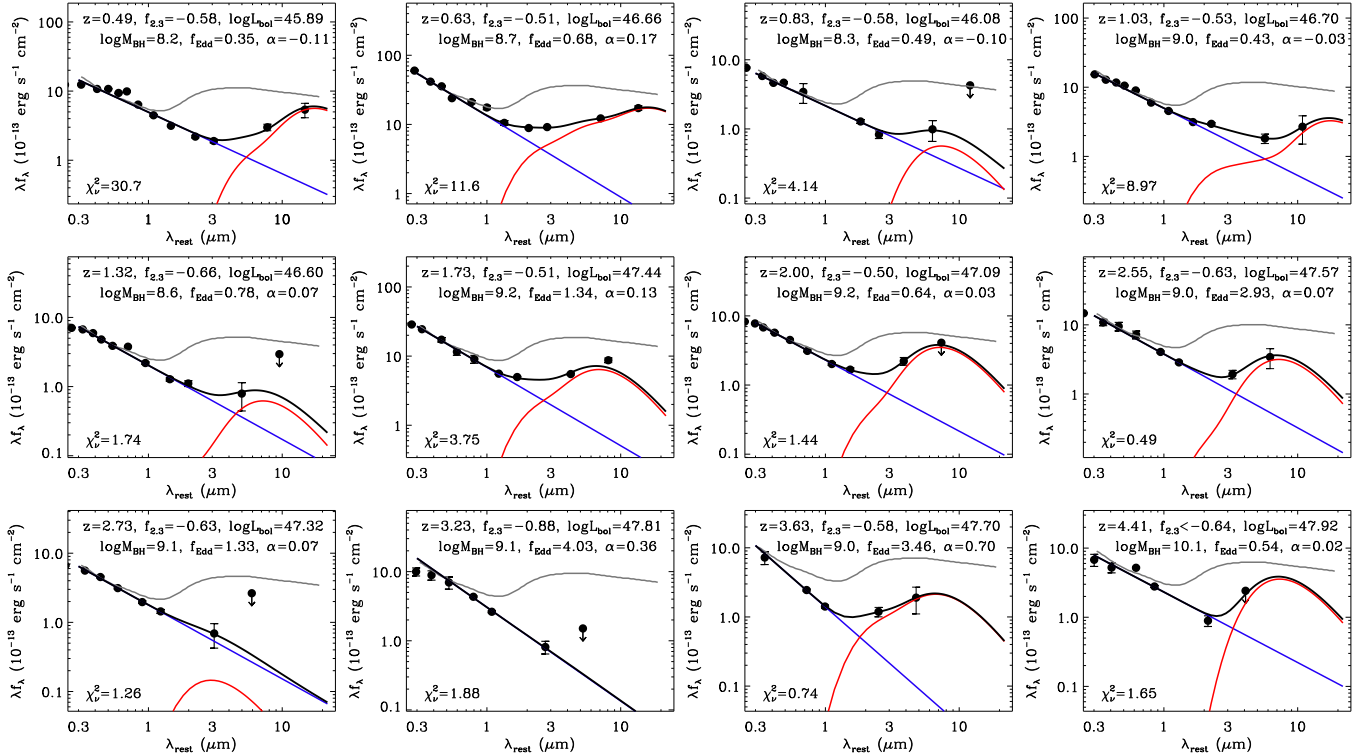}
\caption{Selected list of hot dust poor quasar SEDs, sorted in redshift. Along with the observed data points and 2\,$\sigma$
upper limits, model fits with accretion disk (blue) and dust components of T=1250, 500 and 200\,K phases (red) are overplotted. 
We include the 500\,K and 200\,K black bodies only when the longest wavelength of each SED exceeds 3.5\,$\mu$m and 
9\,$\mu$m respectively, except for calculating upper limits in $f_{2.3}$/$f_{10}$. In addition, the composite SED of luminous ($L_{\mathrm{bol}}\gtrsim10^{46}\,$erg\,s$^{-1}$) SDSS 
quasars (gray, \citealt{Ric06}) normalized to the data at 5100\,\AA, are overplotted. SDSSJ144706.80+212839.3 at $z=3.23$ 
(bottom, 2nd panel from left), is the quasar with the smallest NIR-to-optical flux ratio of $f_{2.3}$\,=\,$-0.88$, reaching 
down to similar values to J10 dust free quasars. \label{fig5}}
\end{figure*}
\begin{deluxetable*}{ccccccc}
\centering
\tablecolumns{7}
%\tabletypesize{\scriptsize}
\tablecaption{Hot dust poor quasar properties}
\tablewidth{0.9\textwidth}
\tablehead{
\colhead{Name} & \colhead{$z$} & \colhead{$f_{2.3}$} & \colhead{$f_{10}$} & \colhead{log\,$L_{\mathrm{bol}}$} & 
\colhead{log\,$M_{\mathrm{BH}}$} & \colhead{$f_{\mathrm{Edd}}$}}
\startdata
J001754.17+011123.4 & 1.465 & -0.77$\,\pm\,$0.04 & 99.00$\,\pm\,$99.00 & 46.503$\,\pm\,$0.020 & 8.47$\,\pm\,$ 0.32 & 0.836$\,\pm\,$0.609\\
J005205.56+003538.2 & 0.399 & -0.52$\,\pm\,$0.01 & -0.48$\,\pm\,$0.03 & 46.521$\,\pm\,$0.004 & 8.43$\,\pm\,$ 0.01 & 0.959$\,\pm\,$0.030\\
J012112.14--003037.1 & 1.654 & -0.64$\,\pm\,$0.03 & 99.00$\,\pm\,$99.00 & 46.594$\,\pm\,$0.020 & 8.75$\,\pm\,$ 0.07 & 0.532$\,\pm\,$0.088\\
J013113.42--093401.1 & 1.149 & -0.53$\,\pm\,$0.08 & -0.43$\,\pm\,$0.18 & 46.759$\,\pm\,$0.027 & 9.18$\,\pm\,$ 0.08 & 0.293$\,\pm\,$0.058\\
J014036.47+000335.9 & 1.636 & -0.58$\,\pm\,$0.02 & 99.00$\,\pm\,$99.00 & 46.975$\,\pm\,$0.011 & 10.09$\,\pm\,$ 0.03 & 0.059$\,\pm\,$0.005
\enddata
\tablecomments{Catalog of hot dust poor quasars sorted in RA, where only the first five rows are displayed. The units for 
$L_{\mathrm{bol}}$ and $M_{\mathrm{BH}}$ are erg s$^{-1}$ and $M_{\odot}$. Empty parameters are fixed to 99. The entire 
table is published in the electronic edition of the paper. A portion is shown here for guidance regarding its form and content.}
\end{deluxetable*}

For the MgII fitting, we note that the DR7 and DR9 spectra are treated differently in S11, such that the DR7 broad MgII 
FWHMs are measured with and without subtracting the narrow ($<$1200 km s$^{-1}$) component, while the narrow line 
component itself is not used for the DR9 fitting. In order to shift each MgII mass estimator to be mutually consistent, 
we followed the argument of S11 to normalize each MgII $M_{\mathrm{BH}}$ recipe to the H$\beta$ $M_{\mathrm{BH}}$ of 
\citet{Ves06}, obtaining the coefficient $a$ in Equation 3 while keeping $b$ and $c$ fixed. This process was performed 
for all 2489 objects covering both H$\beta$ and MgII emission lines at 0.7\,$<$\,$z$\,$<$\,0.8 with continuum S/N\,$>$\,5. 
We find $a$\,=\,6.75 when using the MgII line width measured subtracting the narrow component, consistent with 6.74 
in S11, and $a$\,=\,6.81 when the narrow component is included. Next, we normalized the DR9 MgII $M_{\mathrm{BH}}$ with 
that of 1125 overlapping DR7 objects at 0.7\,$<$\,$z$\,$<$\,2.1 to find $a$\,=\,6.79, irrespective of whether the narrow 
MgII lines of DR7 quasars are subtracted or not, and even when these two DR7 masses are averaged. Since it is still 
debatable whether to subtract the narrow component for the MgII line width measurement (e.g., S11), we averaged the narrow 
component subtracted/included $M_{\mathrm{BH}}$ for the DR7 sample giving the least scatter with the overlapping DR9 
$M_{\mathrm{BH}}$, but at the same time caution on the accuracy of S11 MgII $M_{\mathrm{BH}}$ until the fitting dependent 
systematic uncertainties are clarified in the future. Lastly, we substituted 7 MgII $M_{\mathrm{BH}}$s at $z<$\,2.1 with 
low S/N, with that from CIV. 

In addition to the black hole mass estimates, we computed the bolometric luminosities from the 5100\,\AA\, monochromatic 
luminosity, $L_{5100}$, which is derived from our own broad-band SED fit. This approach may reduce the systematic uncertainties from 
extinction or BAL features, compared to when converting the rest-frame UV into bolometric luminosity. We used a constant 
bolometric correction of 9.26 (S11), derived from the composite quasar SED in \citet{Ric06}. To check whether our photometrically 
determined 5100\,\AA\, luminosities are reliable, we compared our values with that from the spectral fitting in S11, to find 
a good agreement with a difference of $L_{5100}$\,$-$\,$L_{5100,S11}$\,=\,0.06\,$\pm$\,0.16 dex.

\section{Results}
\subsection{Number counts and SEDs of dust poor quasars}
\begin{figure*}
\centering
\includegraphics[scale=0.9]{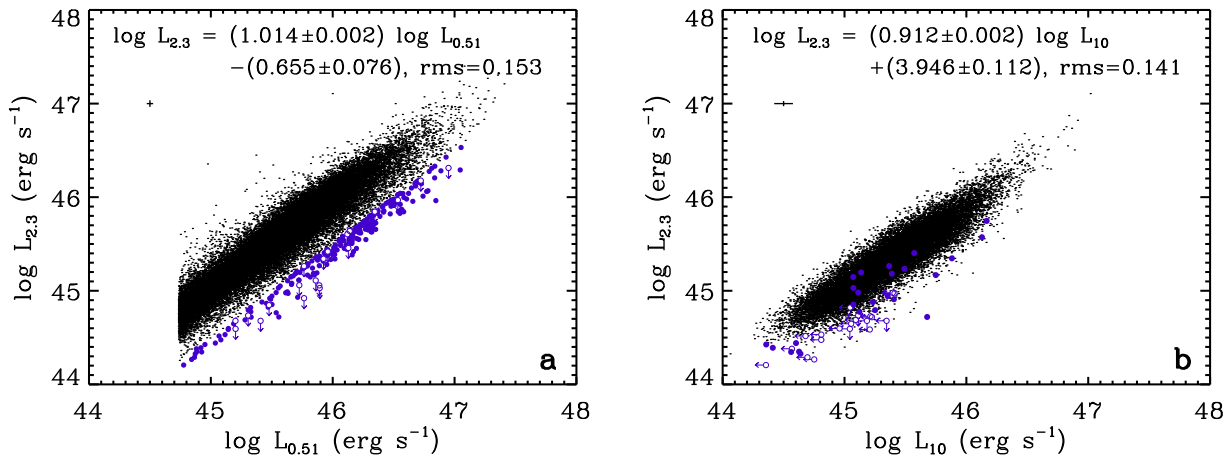} 
\caption{Correlations between optical and NIR monochromatic luminosities (left), and between MIR and NIR luminosities (right). 
The distributions of hot dust poor quasars (purple, filled circles represent detections and open arrows are based on WISE 
2\,$\sigma$ upper limits at corresponding wavelengths) and of the rest of the sample (black dots, detections only) are plotted, 
together with an overall median error bar on the upper left of each panel. Linear fits to luminosity correlations are obtained 
by applying the ordinary least squares bisector method \citep{Iso90} to the entire sample with detections, where the fitting 
coefficients and rms scatter are displayed. \label{fig6}}
\end{figure*}

There are 253 objects that meet the hot dust poor quasar criterion $f_{2.3}$\,$<$\,$-0.5$. Visually inspecting individual 
images and spectra we find 17 blended images in SDSS, and 3 stellar spectra that are rejected. After all, we keep 233 hot 
dust poor quasars out of a total of 40,805, which is 0.6\% in fraction. Figure 4 shows the composite SED of hot dust poor 
quasars together with that of the whole sample. Both composites were constructed by the geometric mean as to preserve 
the global continuum shape \citep{Van01}, after normalizing individual SEDs at 0.51\,$\mu$m. We fitted the composite SEDs 
by 3rd order polynomials in the UV ($<$\,0.3\,$\mu$m) and with optical power-law plus IR black body combinations (\S3), 
to obtain the overall shape information, and to compare the bolometric corrections forthcoming (\S4.2). The composite hot 
dust poor SED in Figure 4 is bluer in the optical, and about three times fainter in the rest-frame 2.3\,$\mu$m than the 
composite of all SEDs. 

Individual hot dust poor SEDs overplotted to their composite in Figure 4, are also depicted with model fits in Figure 5, while 
their properties are summarized in Table 2. We find that the majority of hot dust poor quasars exhibit warm dust emission 
under the detection of NIR--MIR, even if the hot dust component is negligible. Exceptions are objects for which upper limits 
in the rest-frame MIR do not allow us, to assess whether they are lacking the warm dust component or not. Specifically, the 
optical--NIR spectral shapes of the upper $W$4 limit objects at $z$\,$>$\,2 in Figure 5, are similar to the two dust free 
quasars in J10. Nevertheless, we would like to be careful to call these objects ``dust free'', where we instead use the term 
``dust poor'' until deep MIR data are available to tell the absence of warm dust emission. Although mostly present in warm 
dust emission, our hot dust poor quasars are in general very weak in hot dust emission, belonging to a small subset of hot 
dust poor AGN selection in other studies. Our hot dust poor quasars fall within the classification of Class II hot dust poor 
quasars in H10 that are power-law shaped in the optical--NIR, or optically blue hot dust poor quasars in M11. 

\begin{figure}
\centering
\includegraphics[scale=0.9]{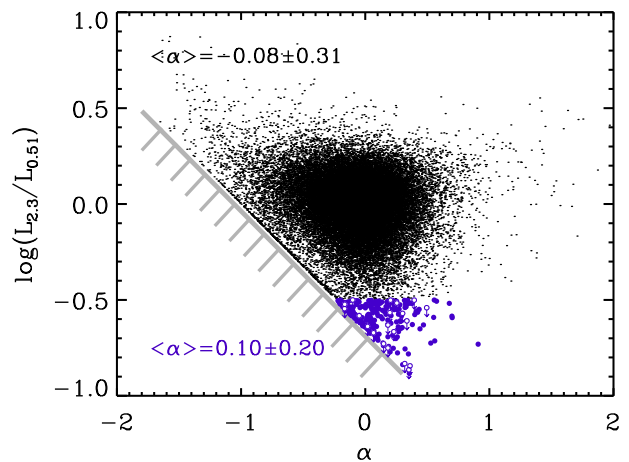}
\caption{Distribution of $f_{2.3}$ along the continuum slope $\alpha$ (left), following the color and symbol layouts of 
Figure 6. The average and standard deviation of $\alpha$ are displayed for both the entire and dust poor samples. The 
data points forming the diagonal line just above the gray highlighted line, are SEDs fitted as purely power-law in the 
optical--NIR, with no hot dust component contribution ($c_{hd}$\,=\,0). The gray shaded region are prohibited since they 
require $c_{hd}$\,$<$\,0. \label{fig7}}
\end{figure}
\begin{figure*}
\begin{center}$
\begin{array}{cc}
\includegraphics[scale=0.9]{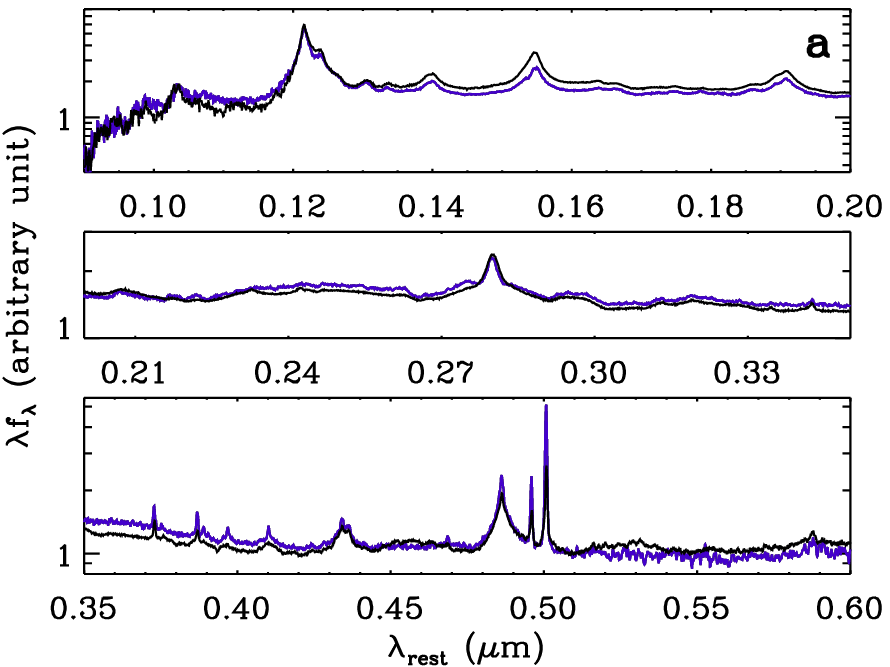} &
\includegraphics[scale=0.9]{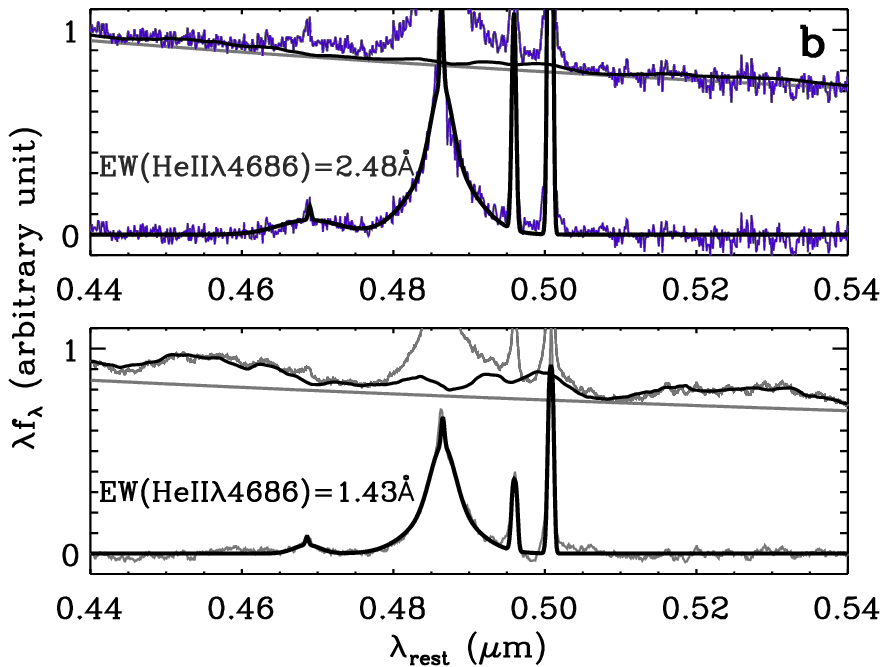} 
\end{array}$
\end{center}
\caption{Left: Composite spectra of dust poor quasars (purple) and of general SDSS quasars (black, from \citealt{Van01}), 
smoothed by 3 pixels to display the spectral features better. The number of spectra used to construct the composite from 
the top to bottom panels are 163, 125, 22 for dust poor quasars, and around 500, 1800, 700 from \citet{Van01}. Right: 
Spectral fits around the H$\beta$ region for dust poor (top) and ordinary (bottom) composites. The measured 
HeII$\lambda$4686 equivalent widths are printed for both samples. \label{fig8}}
\end{figure*}

To better understand the photometric properties of hot dust poor quasars, we plot the luminosity correlations in Figure 6. 
The NIR versus optical luminosities in Figure 6a shows that $L_{2.3}$ is in general proportional to $L_{0.51}$, meaning luminous 
AGNs show stronger dust emission than faint AGNs (e.g., \citealt{Haa03}). With respect to this relation, we find that hot 
dust poor quasars are objects defining the lower $\sim$\,3.2\,$\sigma$ envelope. Next, warm and hot dust emission 
strengths are compared through the 2.3\,$\mu$m versus 10\,$\mu$m luminosities in Figure 6b, which shows a linear relation 
between the MIR--NIR luminosities. Interestingly, the detected hot dust poor data points lie moderately below the relation 
by 1.8\,$\sigma$ on average, suggesting perhaps a shifted MIR--NIR luminosity relation for hot dust poor SEDs. Nonetheless, 
we may find hot dust poor quasars to stay on the warm dust poor side, as their 10\,$\mu$m luminosities are lower by 
1.9\,$\sigma$ on average to the linear relation, $\log(L_{10})=(1.148\pm0.004)\log(L_{0.51})-(6.627\pm0.188), 
\mathrm{rms}=0.200$, derived from the entire sample with 10\,$\mu$m detections. 
Therefore we find it reasonable from our NIR/MIR luminosity correlation of hot dust poor quasars, although with a possible 
offset, to deduce that hot dust emission is a marginally good tracer of warm dust emission in AGNs. The IR portion of the quasar 
SED to peak in the MIR (e.g., \citealt{Ric06}), and the acceptable correlation between MIR/far-IR (hereafter FIR) luminosities 
of quasars (e.g., \citealt{Haa03}), further suggest hot dust poor sources to stay relatively dust poor through the entire IR. 
For these reasons, we choose to refer to ``hot dust poor'' quasars as ``dust poor'' quasars from now on. 

To investigate how dust poor quasars differ from ordinary quasars in wavelengths other than the NIR/MIR, we plot 
$f_{2.3}$ as a function of optical continuum slope $\alpha$, in Figure 7. The average slope and its scatter of 
$<$$\alpha$$>$\,=\,$-0.08$\,$\pm$\,0.31 from our entire sample are ranged somewhat bluer than $-0.44$ of \citet{Van01}. 
This is consistent with the results of \citet{Dav07} where more luminous quasars have bluer continuum slopes, as our sample 
are high luminosity selected SDSS quasars. We also note that differences in the fitting range or the method to measure 
$\alpha$, could further shift the values of $\alpha$ (e.g., Table 5 in \citealt{Van01}). In any case 
$<$$\alpha$$>$\,=\,0.10\,$\pm$\,0.20 of dust poor quasars, comes roughly close to $\alpha=\frac{1}{3}$ of optically thick 
and locally heated accretion disk model predictions, and of polarized observations of quasar SEDs seen through the dust 
\citep{Kis08}. This indicates that the optical continuum from the accretion disk of dust poor quasars are mostly unobscured 
to our line of sight, consistent with the weak IR re-emission observed. Meanwhile, there are points forming a diagonal 
line across the lower left part of $\alpha$--$f_{2.3}$ space in Figure 7, which are optical to NIR SEDs best fit by the 
continuum component only. Although these objects do not seem to involve the hot dust component for the SED fitting, 
satisfying $c_{hd}$\,=\,0 in Equation 2, the NIR emission could be taken over by the extrapolated optical continuum component, 
especially at the red end of small $\alpha$. Therefore we regard at least part of the diagonal sequence in Figure 7, to 
effectively mean lower limits in the measured $\alpha$.

\begin{figure}
\centering
\includegraphics[scale=0.9]{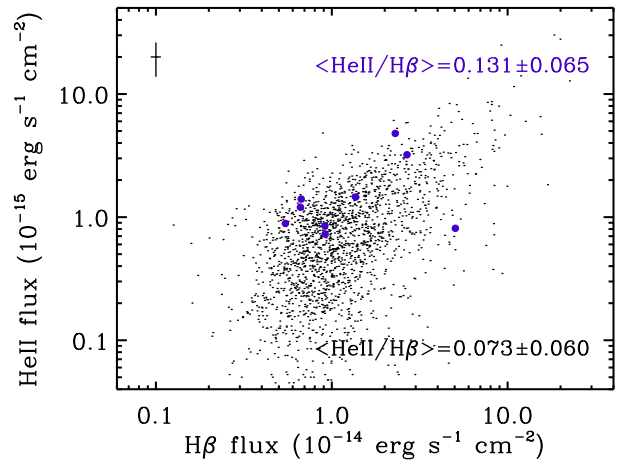}
\caption{Distribution of HeII$\lambda$4686 to H$\beta$ broad line fluxes, following the color and symbol layouts of Figure 
6. The H$\beta$ and HeII$\lambda$4686 line fluxes are measured for objects at $z$\,$<$\,0.8, with an overall median error 
bar drawn on the upper left. Out of 3880 continuum S/N\,$>$\,10 spectra 1899 have non-detections in H$\beta$ or HeII, 
or fail to meet H$\beta$ line S/N\,$>$\,2, where they are excluded from the plot. The average and standard deviation of 
HeII/H$\beta$ on the plot, are printed for both the entire and dust poor samples. \label{fig9}}
\end{figure}
\begin{figure*}
\centering
\includegraphics[scale=0.9]{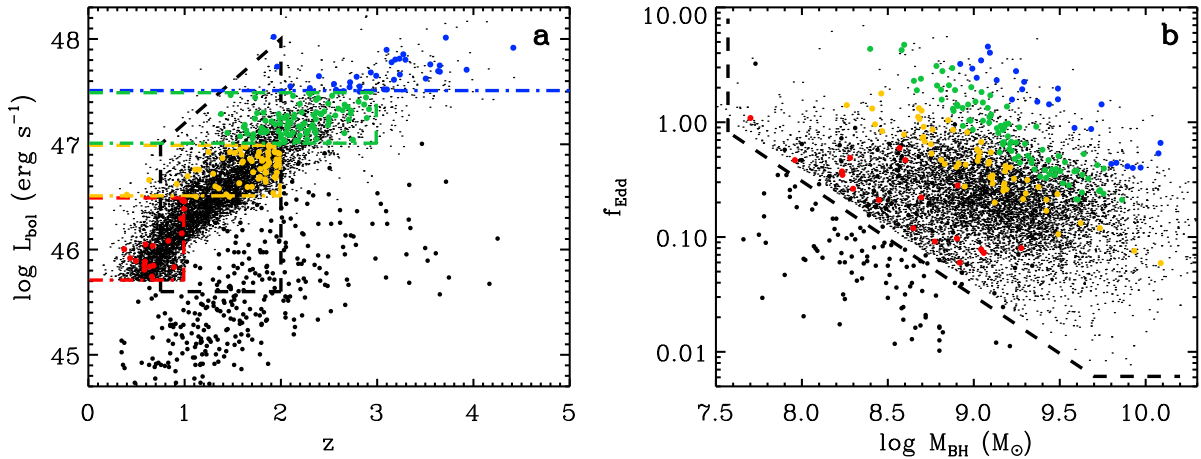}
\caption{Redshift--bolometric luminosity (left) and black hole mass--Eddington ratio distribution (right) of our sample. 
The range of data points from M11 are highlighted (black dashed lines) for comparison. Furthermore, the 361 
\textit{XMM}-COSMOS spectroscopically confirmed type-1 AGNs in \citet{Lus10} that are similar to the sample of H10, are 
overplotted (black circles). Since the entire sample is large in number, a quarter of the sample are randomly selected 
and plotted (black dots) in both panels, to better resolve and compare the overall distribution to that of dust poor objects 
(color circles). We subdivide our sample into four bolometric luminosity bins (color dot dashed lines), for the forthcoming 
analysis to be independent of luminosity selection effects. \label{fig10}}
\end{figure*}

In addition to the optical continuum slope measurements from photometric data points, we constructed a composite 
UV--optical spectrum of dust poor quasars plotted in Figure 8a, along with the composite of general SDSS quasars in 
\citet{Van01}. Compared to typical quasars, dust poor quasars have a bluer NUV--optical continuum slope. Moreover, fitting 
all prominent UV to optical emission lines we find the HeII$\lambda$4686 line emission of dust poor quasars to be especially 
stronger, where the equivalent widths are about 1.7 times than that measured using the composite from \citet{Van01}, in 
Figure 8b. For the HeII fitting we subtracted the continuum and broad FeII in a similar way as S11 but with the template 
of \citet{Tsu06}, while a single Gaussian profile was used for both broad and narrow emission components of HeII. The HeII 
region was simultaneously fitted with H$\beta$ and [OIII] lines to better decompose the HeII line emission. 

Because the HeII$\lambda$4686 line is relatively well decoupled from neighbor emission lines, and to check whether the 
strong HeII of the dust poor composite spectrum are reflected in the individual spectra, we further calculated 
the HeII/H$\beta$ line ratios for all $z$\,$<$\,0.8 DR7 objects with spectral continuum S/N\,$>$\,10 and H$\beta$ line 
S/N\,$>$\,2. The results are plotted in Figure 9, where HeII and H$\beta$ line fluxes of dust poor quasars are compared 
with those of ordinary sources. Dust poor quasars occupy regions with about 1.8 times higher HeII/H$\beta$ line ratios 
compared to typical quasars, being on average 13\% and 7\% respectively, consistent with the results from Figure 8b. This 
can be understood as a result of the strong UV continuum source, as the HeII/H$\beta$ line ratio and the source spectral 
shape in the extreme-UV (EUV) form a relation \citep{Pen78} as
\begin{equation} \frac{L_{\mathrm{HeII}\lambda4686}}{L_{\mathrm{H}\beta}}=1.99\times4^{\alpha_{EUV}}, \end{equation}
where $\alpha_{EUV}$ indicates the EUV slope of the source that brings HI and HeII emission. Equation 4 yields a bluer slope of 
$\alpha_{EUV}$\,$\sim$\,$-2.0$ for dust poor quasars, which hints their broad line regions to be more showered by energetic 
UV photons than average, consistent with the observed bluer EUV spectra in the left part of Figure 8a top panel. This, 
together with the blue NUV--optical spectra of dust poor quasars in Figure 8a, suggests high temperature, energetic 
radiation sources to be connected to the dust poor property. An example of such a case would be the model prediction of 
accretion disk temperature dependent UV--optical color of AGNs (\citealt{Bon07}).

\subsection{Parameter space study of dust poor quasars}
Figure 10 shows the distribution of AGNs in our sample in the $L_{\mathrm{bol}}$ versus redshift, and $f_{\mathrm{Edd}}$ 
versus $M_{\mathrm{BH}}$. Also plotted are the range of the parameter space covered by other studies (H10 and M11). Here 
the dashed boundary for the M11 sample is determined following their selection criteria, and we mimic the distribution 
of the H10 sample by plotting with thick black points the objects in \citet{Lus10} sample which contain 88\% of the 408 AGNs 
in H10. Overall, our sample are brighter than of H10 in $L_{\mathrm{bol}}$, and less extensive in $L_{\mathrm{bol}}$ but 
more extensive in redshift space coverage than of M11. From Figure 10, we find that the $L_{\mathrm{bol}}$ and $i$-band 
limits (\S2) introduce the luminosity cut at faint end which is variable with redshift. Moreover, the anti-correlation between 
$L_{\mathrm{bol}}$ and the hot dust covering factor CF$_{hd}$ (e.g., M11), may act as a luminosity selection effect (\S5.1) 
to mix up the distributions of $f_{2.3}$ or CF$_{hd}$ when plotted against a parameter closely associated with $L_{\mathrm{bol}}$, 
such as $M_{\mathrm{BH}}$ or $f_{\mathrm{Edd}}$. To minimize the luminosity selection effect, we split the whole sample 
into four volume-limited subsamples where 204 dust poor quasars remain inside the dot dashed limits in Figure 10a. 

Now, we use the fraction of dust poor quasars (hereafter, $p_{hdp}$), to visualize the global trends of dust poor quasars 
in observed parameters. When the $p_{hdp}$ are plotted against $L_{\mathrm{bol}}$ in Figure 11a, we find a positive 
correlation between the two products, in accord with the distribution of $f_{2.3}$ decreasing with $L_{\mathrm{bol}}$ in 
Figure 11c. The first parameter to look into through $p_{hdp}$ under the volume-limits is the redshift, as we would like 
to know the evolution of the fraction of dust poor quasars independent of luminosity selection. In Figures 11b and 11d 
we find not only a trend of lower $f_{2.3}$ at higher redshift overall (bottom panel), but also a clear tendency of increased 
$p_{hdp}$ at higher redshifts at given luminosity (top panel), mixed with the $p_{hdp}$ increasing with luminosity at a 
given redshift. Compiling our $p_{hdp}$ with that inferred from J10, we fit the $z$\,$>$\,2, 
$L_{\mathrm{bol}}$\,$>$\,10$^{47}\,$erg\,s$^{-1}$ data points in Figure 11b to model the redshift evolution of the dust 
poor fraction as $p_{hdp}=(1.35\pm0.13)$\,$\times$\,10$^{-3}$\,$(1+z)^{2.34\pm0.08}$. 

\begin{figure*}
\begin{center}$
\begin{array}{cc}
\includegraphics[scale=0.85]{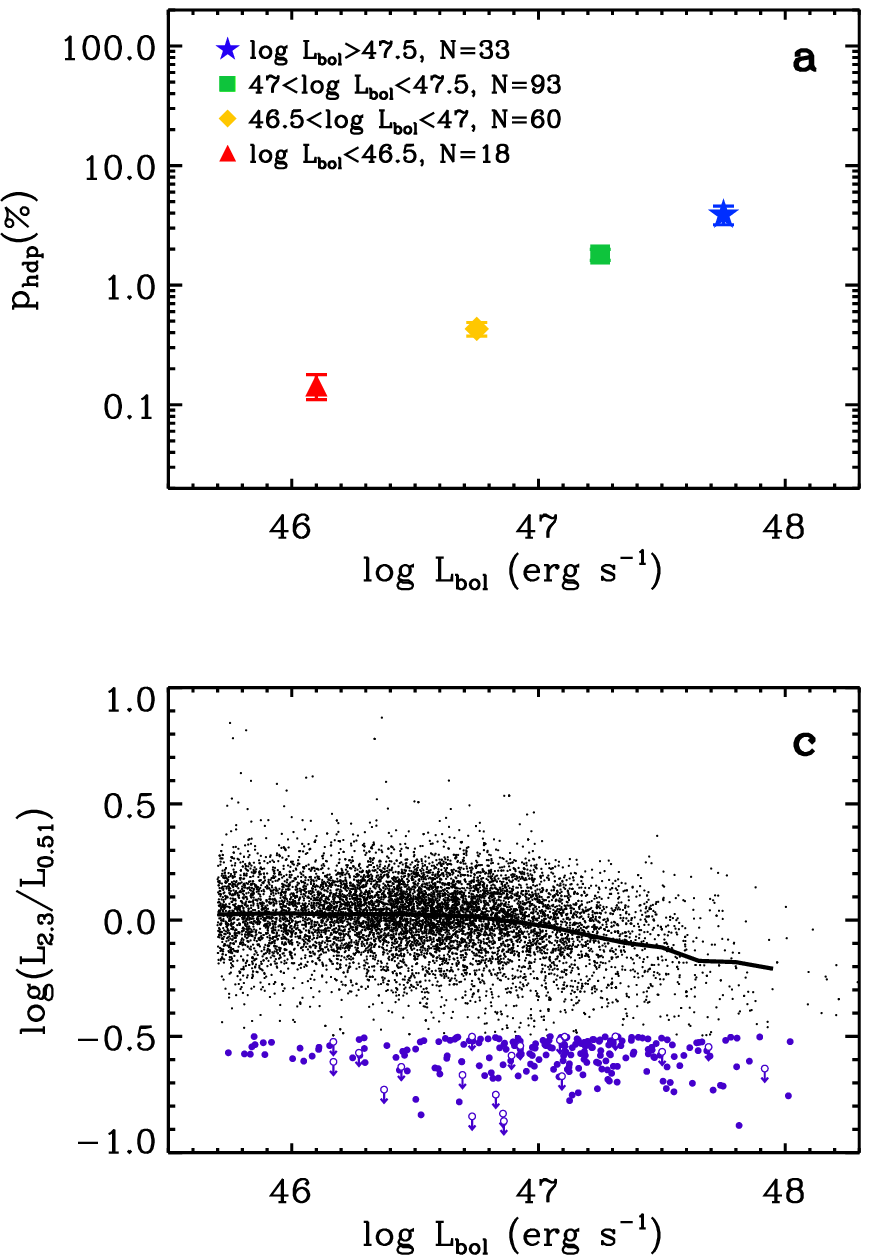} &
\includegraphics[scale=0.85]{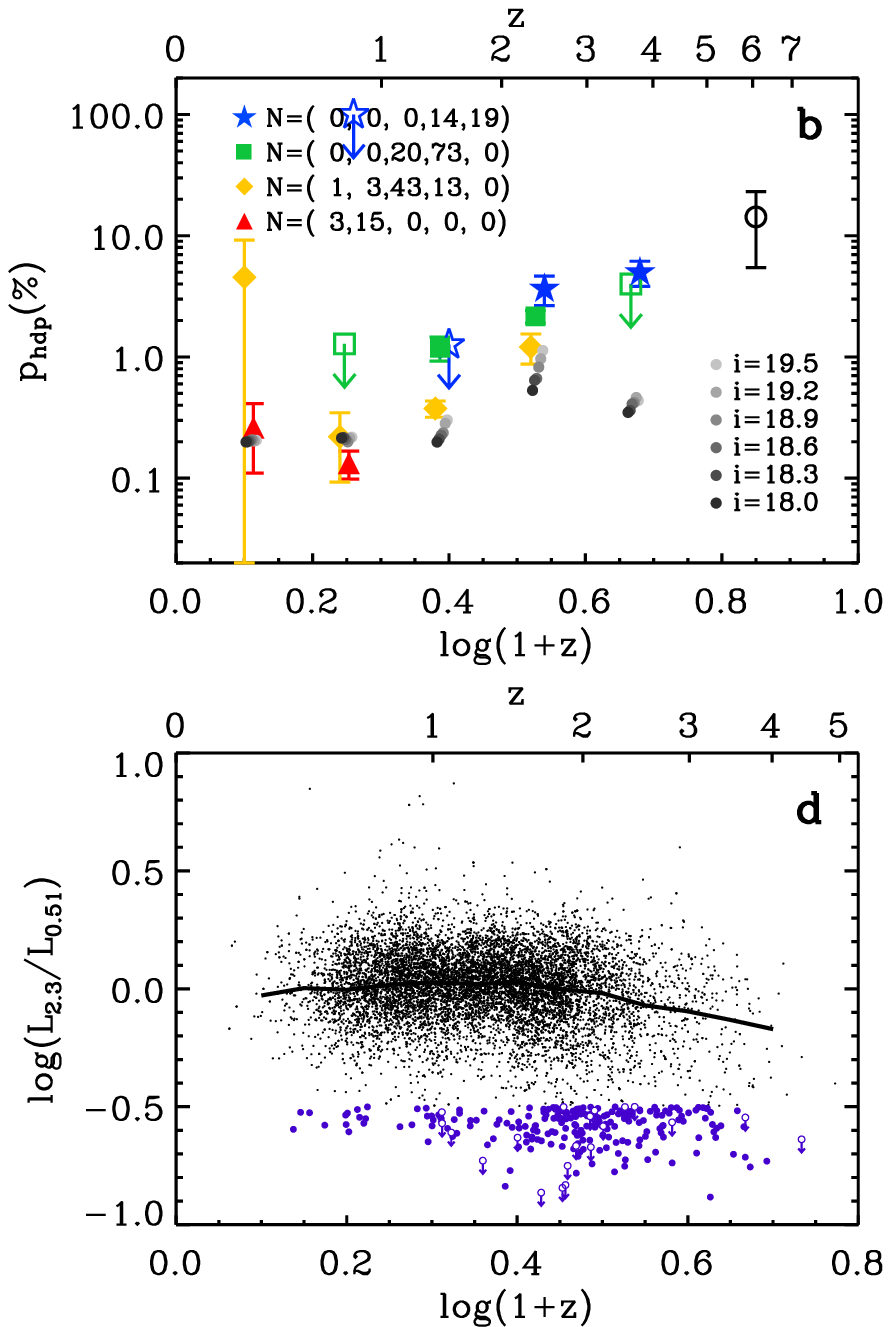}
\end{array}$
\end{center}
\caption{Top left: $L_{\mathrm{bol}}$--hot dust poor fraction ($p_{hdp}$) plot. Color filled symbols represent the fraction 
of quasars in each luminosity bin, to be hot dust poor. The error bars are calculated based on the Poisson number counting 
statistics. Bottom left: $L_{\mathrm{bol}}$--$f_{2.3}$ plot. Individual distribution of hot dust poor quasars (purple) with 
randomly selected one quarter (following Figure 10) of the entire sample (black), are plotted. In addition, the average 
$f_{2.3}$ along $L_{\mathrm{bol}}$ are overplotted in thick solid lines. Top right: $z$--$p_{hdp}$ plot. Color filled 
symbols slightly shifted to each other to avoid overlaps, represent the fraction of quasars in each luminosity and redshift 
bin to be hot dust poor, with each number of objects per data point printed following the arrangement of the data. 
Meanwhile, the open circle is the $z$\,=\,6 result of J10. Upper limits in $p_{hdp}$ are calculated as 1$/N_{bin}$, 
when there are no dust poor quasars in the bin with size $N_{bin}$. The gray gradational circles are test results of 
our $z$\,$<$\,1, $K$ and $W$4 detected sample SEDs, simulated to each redshift and observed $i$-band magnitude. 
Bottom right: $z$--$f_{2.3}$ plot. The layout follows that of the bottom left panel. \label{fig11}}
\end{figure*}

\begin{figure*}
\begin{center}$
\begin{array}{cc}
\includegraphics[scale=0.85]{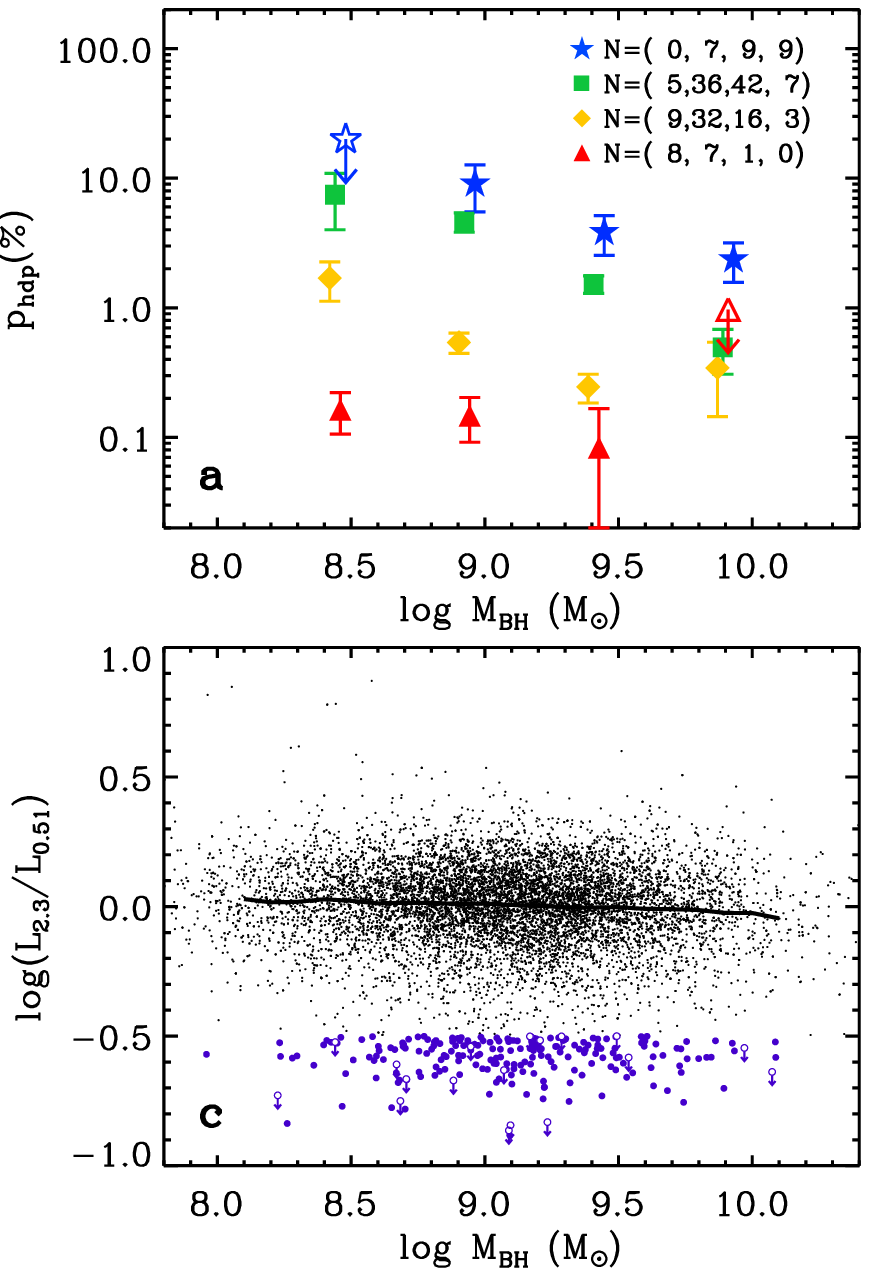} &
\includegraphics[scale=0.85]{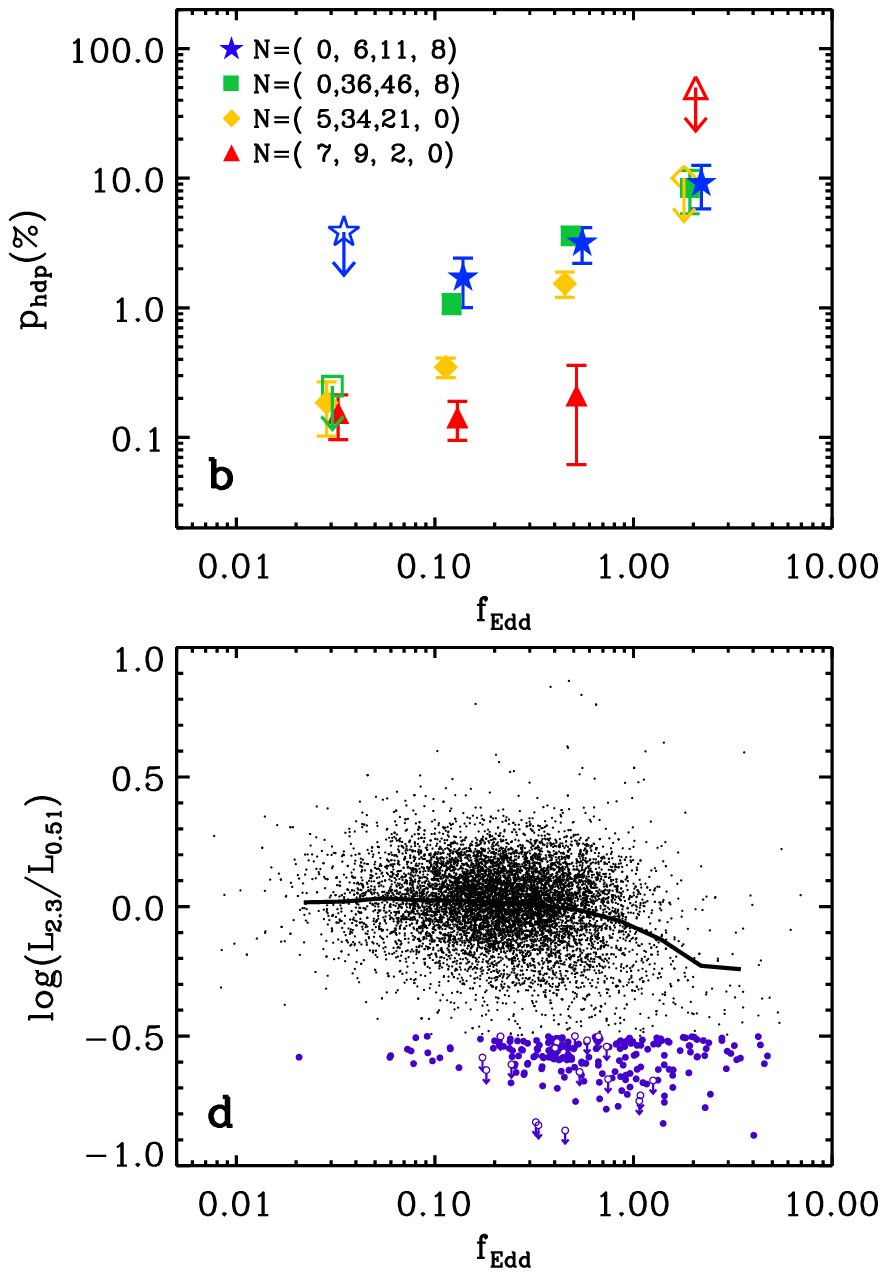}
\end{array}$
\end{center}
\caption{Left: $M_{\mathrm{BH}}$--$p_{hdp}$ plot (top), and the distribution of $M_{\mathrm{BH}}$--$f_{2.3}$ (bottom). 
Right: $f_{\mathrm{Edd}}$--$p_{hdp}$ plot (top), and the distribution of $f_{\mathrm{Edd}}$--$f_{2.3}$ (bottom). The 
meaning of the data point symbols follows that of Figure 11. \label{fig12}}
\end{figure*}

Since we are dealing with a small number of special objects, some of dust poor quasars could be misclassified normal quasars 
due to large photometric errors, or sparse wavelength sampling in the WISE data points. To estimate what fraction of dust 
poor quasars could have arisen from these artifacts, we selected 7834 SDSS DR7 quasars in our sample at $z$\,$<$\,1 with 
$K$ and $W$4 detections, 
and redshifted their SEDs while adding appropriate noises to make them look to have $i$=18--19.5\,mag. Starting 
from a set of observed fluxes and errors linearly interpolated to simulated redshifts ($F_{obs}, \Delta F_{obs}$), we assigned 
the flux errors corresponding to the simulated fluxes as $\Delta F_{sim}\,$=$\,\Delta F_{obs}10^{0.2(m_{sim}-m_{obs})}$ 
where $m_{obs}, m_{sim}$ are observed and simulated magnitudes. Then, we added a Gaussian random noise 
($\sigma_{gauss}\,$=$\,\Delta F_{sim}$) to the simulated fluxes, to obtain  a set of fluxes and errors ($F_{sim}, 
\Delta F_{sim}$). Every object was repeatedly simulated 50 times with randomly added errors, for a given simulated 
redshift and magnitude. Finally the redshifted, error scaled, and noise added set of SEDs were fitted to calculate the dust 
poor fraction. The dust poor fractions from this simulation are overplotted in Figure 11b with gray circles. 

The figure indicates that the objects selected for this test are representative of general quasars at $z$\,$<$\,1, since the 
simulated $p_{hdp}$ regardless of the test magnitude, are within errors to the observed $p_{hdp}$ at $z$\,$<$\,1. But then, 
the test gives the possibility of dust rich ($f_{2.3}$\,$>$\,$-0.5$) quasars to be artificially classified as dust poor, as the 
simulated $p_{hdp}$ starts to increase at $z$\,$>$\,1. Not only does $p_{hdp}$ increase for fainter simulated magnitudes, it 
also has a peak at 2\,$<$\,$z$\,$<$\,3 where the rest-frame 2.3\,$\mu$m lies in between the widely separated $W$2$/W$3 bands. 
This indicates that the dust poor selection becomes uncertain when the photometric data points around the peak of the hot 
dust radiation are sparse. Comparing the observed versus the simulated $p_{hdp}$ of $L_{\mathrm{bol}}$\,=\,10$^{46.5-47}\,$erg\,s$^{-1}$ 
sources in Figure 11b for example (yellow and gray points), we estimate $\sim$\,60\% of the faintest objects ($i$\,=\,18.5--19, 
see Figure 2b) in the dust poor sample to be possibly misclassified normal or border-line quasars at 2\,$<$\,$z$\,$<$\,3. 
However, at $z$\,$>$\,3 only a small fraction of dust poor quasars are possible misclassifications, suggesting that the 
observed $p_{hdp}$ are genuinely higher at $z$\,$>$\,3. Furthermore, taking into account the higher possibility of identifying 
false dust poor objects at 2\,$<$\,$z$\,$<$\,3, the intrinsic distribution of true dust poor quasars will show a stronger 
positive $z$--$p_{hdp}$ relation at $z$\,$>$\,2, for any given luminosity bin. Therefore we regard the overall trend of 
increased $p_{hdp}$ with redshift to remain unchanged under possible artifacts. 

Next, $p_{hdp}$ versus black hole mass and Eddington ratio are plotted in Figures 12a and 12b for four different 
$L_{\mathrm{bol}}$ bins. We find higher $p_{hdp}$ for more luminous $L_{\mathrm{bol}}$ binned subsamples, which is 
consistent with Figures 11a and 11b, and with predictions of smaller dust covering factors at brighter $L_{\mathrm{bol}}$, 
such as from the receding dust torus model \citep{Law91}. Therefore, we note the importance of controlling the range of 
$L_{\mathrm{bol}}$ in order to accurately trace the properties of dust poor quasars. In addition, because our dust poor 
quasars are systematically smaller in CF$_{hd}$, or equivalently larger in NIR bolometric corrections (hereafter BC, 
BC$_{\lambda}$=BC$_{\lambda(\mu m)}$) than average, we would like to double check whether the $L_{\mathrm{bol}}$--$p_{hdp}$ 
trend is affected by a systematically different BC for dust poor SEDs. Utilizing Figure 4 we integrated the model fit on 
the composite photometric SED of all quasars in the 0.04--20\,$\mu$m, to find $L_{0.04-20}/L_{0.51}$\,=\,$6.46$, 
or 70\% of the bolometric luminosity to be bounded within the selected UV through MIR wavelengths, adopting BC$_{0.51}$\,=\,9.26 
from S11. Now, assuming that the SED of the average and dust poor composites do not systematically vary with 
each other in the $\gamma$-ray--X-ray or FIR--radio (see \S5.1 and Table 3 for the assumption to roughly hold for X-ray 
and radio wavelengths), $[(L_{<0.04}+L_{>20})/L_{0.51}]_{hdp}=2.80=(L_{<0.04}+L_{>20})/L_{0.51}$. From our measurement of 
$(L_{0.04-20}/L_{0.51})_{hdp}$\,=\,5.04 out of the composite dust poor SED, we get BC$_{0.51,hdp}$\,=\,7.84, meaning that 
bolometric luminosities of dust poor quasars are about 18\% overestimated with respect to ordinary quasars, under the adoption 
of a single BC$_{0.51}$\,=\,9.26.

Drawing Figures 11--12 again with a 18\% smaller BC$_{0.51}$ for dust poor quasars, however, does not change the strength 
of $p_{hdp}$ trends found in ($z$, $L_{\mathrm{bol}}$, $M_{\mathrm{BH}}$, $f_{\mathrm{Edd}}$) spaces. Therefore we conclude 
that possible systematic biases in the bolometric correction for dust poor quasars does not artificially produce the 
$L_{\mathrm{bol}}$--$p_{hdp}$ relation nor change the observed $p_{hdp}$ trends within the main observed parameters, 
although our test brings caution when applying the BC to dust poor quasars from the monochromatic luminosity alone. After 
all, reading Figures 12a and 12b within each luminosity bin decoupled from the $L_{\mathrm{bol}}$ dependence on $p_{hdp}$, 
we find more dust poor quasars at lower black hole masses and high Eddington ratios. 

\begin{figure*}
\centering
\begin{center}$
\begin{array}{cc}
\includegraphics[scale=0.85]{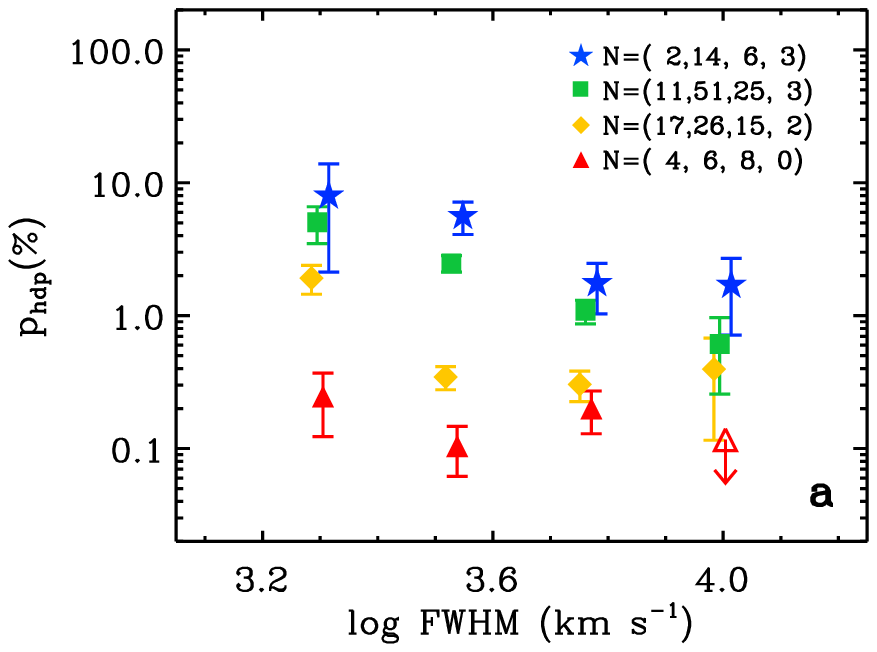} &
\includegraphics[scale=0.85]{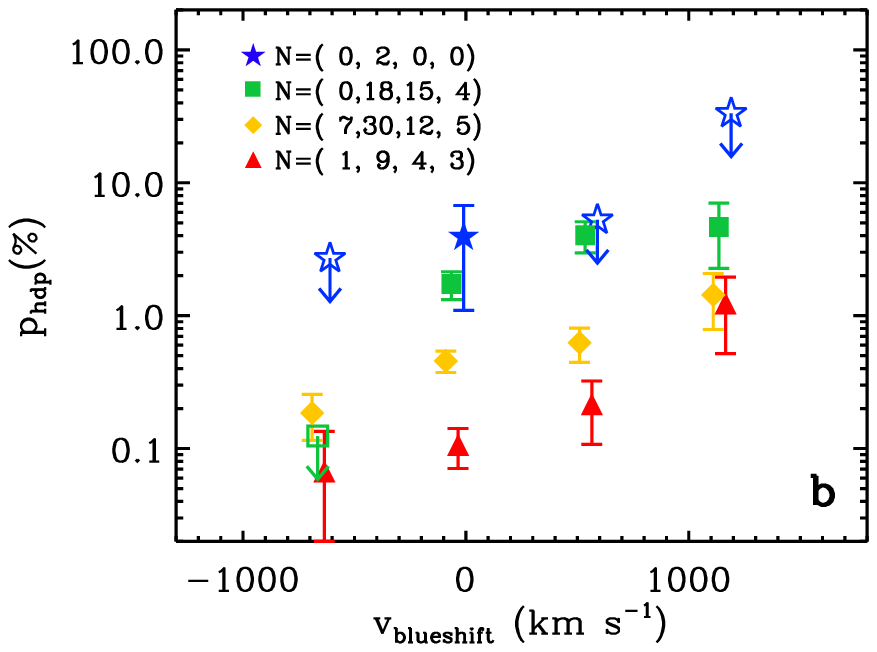}\\
\includegraphics[scale=0.85]{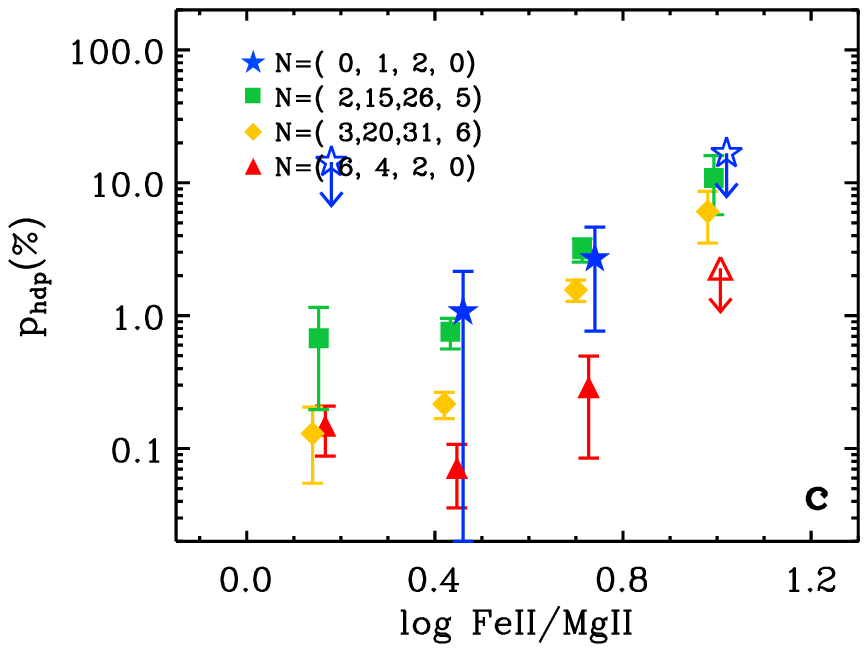} &
\includegraphics[scale=0.85]{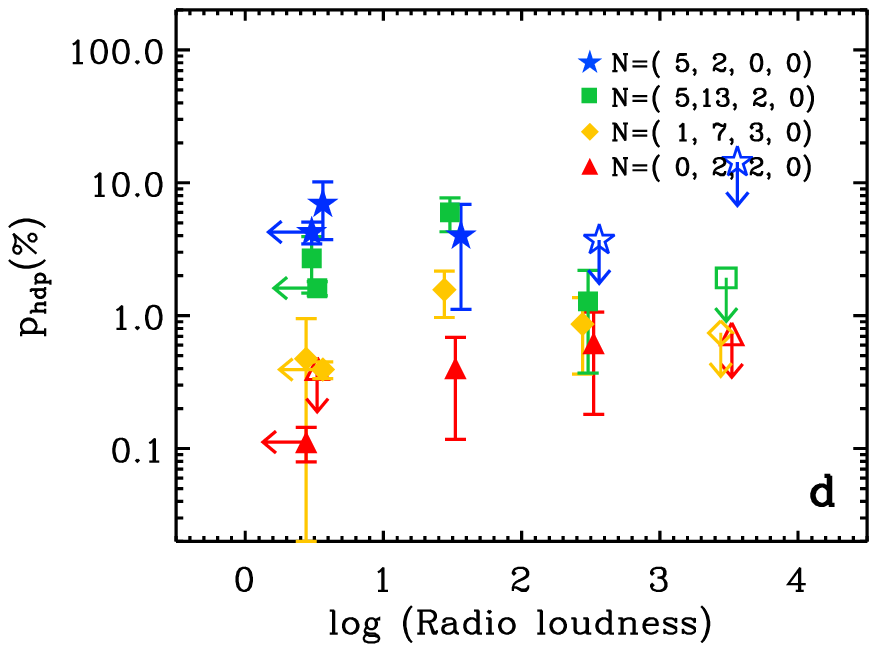}\\
\end{array}$
\end{center}
\caption{$p_{hdp}$ in parameters related to broad emission lines and radio properties. In panels a--d we plot $p_{hdp}$ 
versus broad line FWHM, broad line blueshift from the systemic redshift, FeII(2200--3090\,\AA)/MgII$\lambda$2798 metallicity, 
and radio loudness $R_{radio}=F_{\nu,6cm}/F_{\nu,2500}$. The meaning of the data point symbols follows that of Figure 
11 except for the left pointing arrows in panel d, where they are $p_{hdp}$ estimated without radio detections, but with a 
rough $R_{radio}$\,$\lesssim$\,10 limit. \label{fig13}}
\end{figure*}

Finally, we plot $p_{hdp}$ against other AGN observables in Figure 13, starting from the broad line width and velocity offset. 
We used the FWHM from S11 as broad line widths, updated and replaced for some dust poor quasars in \S3. Meanwhile, we 
took the broad line velocity offset measurements from S11, defined as the relative shift of the broad line center with respect 
to the systemic redshift. The choice of broad line for the line width and offset measurements at each redshift, was identical 
as to measure $M_{\mathrm{BH}}$ (\S3), except that CIV line velocity shifts were not used due to their overall blueshift (S11). There 
is a negative correlation between $p_{hdp}$ and broad line FWHM in Figure 13a, consistent with the trend found for 
$M_{\mathrm{BH}}$, and a meaningful tendency for velocity offset towards blueshifted broad emission, in Figure 13b. The 
implication of these results will be discussed in the next section. Next, reading Figure 13c, dust poor quasars with respect 
to the FeII(2200--3090\,\AA)/MgII$\lambda$2798 metallicity derived from S11, tend to have more iron rich broad line 
regions. Lastly, in Figure 13d we investigated the radio loudness dependence on $p_{hdp}$, with $R_{radio}=F_{\nu,6\mathrm{cm}}/F_{\nu,2500}$ 
from S11. The radio data are from the FIRST survey \citep{Whi97} roughly coinciding with the SDSS coverage and reaching 
a typical 5\,$\sigma$ sensitivity of 1\,mJy. This is similar to the WISE $W$3 sensitivity when assuming a flat source SED in 
$F_{\nu}$ from the MIR to radio, implying that most radio loud/intermediate sources are able to be radio detected. However, 
the FIRST detection threshold (5\,$\sigma$) is higher than that of WISE (2\,$\sigma$), and misses the majority of radio quiet sources, 
which results in only 10\% of the final AGN sample used in this work to be detected. But still, the radio loudness is roughly 
complete down to $R_{radio}$\,$\sim$\,10 within the detection limit, while the distribution of $R_{radio}$ shows a trend 
expected to fall below $R_{radio}$\,$\sim$\,10 under nondetections, for both dust poor and entire AGNs. To further 
investigate the completeness of the radio data, we calculated the $p_{hdp}$ in Figure 13d while adding the FIRST undetections 
into the $R_{radio}$\,$<$\,10 bin, to find the $p_{hdp}$ to be consistent within errors, to that with the detections only.
After all, we find no significant preference of dust poor quasars in the $R_{radio}$--$p_{hdp}$ distribution. Likewise, we 
tried to search for the X-ray loudness features from the RASS observed counts in \citet{Sch10}, but failed to compile enough 
number of matches (N=7) for proper analysis.

\section{Discussion}
\subsection{Observational characteristics of dust poor quasars}
\begin{deluxetable*}{ccccccccc}
\tablecolumns{9}
%\tabletypesize{\scriptsize}
\tablecaption{K--S test on dust poor quasar fraction in observed parameters}
\tablewidth{.95\textwidth}
\tablehead{
\colhead{log\,$L_{\mathrm{bol}}$\,(erg s$^{-1}$)} & 45.7--46.5 & 46.5--47 & 47--47.5 & $>$47.5 & 45.7--46.5 & 46.5--47 & 47--47.5 & $>$47.5 
\\\colhead{Parameter} & \colhead{$p_{KS}$} & \colhead{$p_{KS}$} & \colhead{$p_{KS}$} & \colhead{$p_{KS}$} & \colhead{$N_{hdp}$} & \colhead{$N_{hdp}$} & \colhead{$N_{hdp}$} & \colhead{$N_{hdp}$}}
\startdata
$z$ & -0.09 & 3.1e-06 & 1.2e-03 & 0.13 & 18 & 60 & 93 & 33\\
$M_{\mathrm{BH}}$ & -0.64 & -1.1e-04 & -7.7e-12 & -3.4e-03 & 18 & 60 & 90 & 25\\
$f_{\mathrm{Edd}}$ & 0.44 & 1.8e-07 & 1.6e-10 & 1.2e-04 & 18 & 60 & 90 & 25\\
FWHM & -0.26 & -3.8e-04 & -7.3e-08 & -8.9e-03 & 18 & 60 & 90 & 25\\
$v_{blueshift}$ & 2.8e-03 & 2.9e-05 & 9.0e-11 &  \,\,\,.....\,\,\,  & 18 & 60 & 40 &  3\\
FeII/MgII & -0.26 & 3.8e-04 & 7.3e-08 &  \,\,\,.....\,\,\,  & 16 & 60 & 50 &  3\\
$R_{radio}$ & 0.76 & 0.64 & 0.16 & -0.05 &  4 & 11 & 20 &  7
\enddata
\tablecomments{$p_{KS}$ is the K--S probability for the distribution of dust poor quasars in each parameter and luminosity 
bin, to be indistinguishable from that of the rest of our whole sample. The configuration of volume-limited luminosity bins 
follow Figure 10a. Positive $p_{KS}$ implies dust poor quasars to have a larger parameter value than average quasars, 
and vice versa for negative signs, while statistically meaningful probabilities set as $|p_{KS}|$\,$<$1\%, are printed in 
exponential format. $N_{hdp}$ is the number of dust poor quasars used in each bin to calculate p$_{KS}$, ordered by the 
same luminosity range as for $p_{KS}$. Blank fields correspond to bins with few quasars to perform a test.}
\end{deluxetable*}
To quantify how distinct dust poor quasars are from typical quasars in various physical properties, we performed a set of 
Kolmogorov--Smirnov (K--S) tests by comparing the distribution of dust poor against average samples. The result 
is summarized in Table 3. Although it becomes uncertain at $L_{\mathrm{bol}}$\,$<$\,10$^{46.5}$\,erg\,s$^{-1}$, 
most of the K--S probabilities at $L_{\mathrm{bol}}$\,$>$\,10$^{46.5}$\,erg\,s$^{-1}$ are smaller than 1\%, indicating 
that dust poor quasars are likely to be drawn from a different population to average luminous quasars. Moreover the directions 
of the inequality between dust poor and average quasar properties in Table 3, are consistent with Figures 11--13. To begin 
summarizing the main results from the table and figures, dust poor quasars are at higher redshifts for a given luminosity, 
with this trend found to be secure at $z$\,$>$\,2. This is consistent with J10 and H10, but not with M11 where they do not 
find an evolution in the dust poor fraction. Next, we find dust poor quasars of relatively lower black hole masses 
in the $10^{8}M_{\odot}$ order and high Eddington ratios of super-Eddington accretion, at a given luminosity, in 
agreement with J10 at $z$\,$>$\,6 but not with H10 or M11 where dust poor quasars are indistinguishable to the average in 
$M_{\mathrm{BH}}$ and $f_{\mathrm{Edd}}$. In addition to these fundamental properties, dust poor quasars show narrower 
broad line FWHM of 2000--3000\,km\,s$^{-1}$, blueshifted line centers in the 10$^{3}$\,km s$^{-1}$ order, and higher 
FeII/MgII line ratios by a few times of factors than average. 

Although we find some of the parameter trends summarized above to agree with previous studies, there are also debatable 
results. To investigate further in between conflicting arguments, we would first like to note the limitation of small number 
statistics. The number of our volume-limited dust poor quasars is N=204, which is enough to precisely look for parameter 
trends as we find many of the K--S probabilities in Figure 3 to be statistically significant. Also, our study fills the 
lower redshift counterpart to J10 as they miss the $z$\,$<$\,6 dust poor population, possibly due to the combination of small 
parent sample size in J10 and the $z$--$p_{hdp}$ trend in this work that predicts lesser dust poor quasars at lower--$z$. 
Meanwhile, although H10 selected dust poor quasars with a relatively high fraction ($\gtrsim$\,10\%), their total 
number of N=41 dust poor quasars is still insufficient to tell whether they are drawn from a different population or not, 
as the K--S probabilities lie ambiguously in between 0 and 1 except with respect to the redshift. Therefore, a larger sample 
size would be required to understand in detail, the dust poor quasars defined in H10. We additionally note that the X-ray 
selection of H10 may bring discrepancies to the dust poor population properties based on our optically selected AGNs. 
However, our $f_{2.3}$\,$<$\,$-0.5$ limit is similarly met for only the Class II criterion of H10, which is merely 15\% 
of the dust poor population in H10. Thus, we mostly regard the high $p_{hdp}$ in H10 as from the different definition of 
being dust poor rather than from the X-ray selection of AGNs to elevate the $p_{hdp}$.

Next, we note the difference of studying CF$_{hd}$ in overall distributions from previous works, and under volume-limited 
subsampling here. For instance, the overall distribution of CF$_{hd}$ is roughly independent of $M_{\mathrm{BH}}$ in J10 
($z$\,$<$\,6), M11, and in Figure 12c. However, in Figure 12a we witness clearly higher $p_{hdp}$ for less massive 
$M_{\mathrm{BH}}$ at a given luminosity of $L_{\mathrm{bol}}$\,$>$\,10$^{46.5}$\,erg\,s$^{-1}$. If the sample had 
a wide dynamical range of $L_{\mathrm{bol}}$, there is a higher chance of the anti-correlation between $L_{\mathrm{bol}}$ 
and CF$_{hd}$ to mix up the distribution of CF$_{hd}$ when plotted against a parameter dependent on $L_{\mathrm{bol}}$. 
In other words, quasars with higher $M_{\mathrm{BH}}$ are more likely to be selected as dust poor objects since 
$M_{\mathrm{BH}} \propto \sqrt{L_{\mathrm{bol}}}$ (e.g., Equation 3 with constant bolometric correction), and because 
there are more dust poor quasars with high $L_{\mathrm{bol}}$ (Figure 11a). Revisiting Figures 12a and 12c, this luminosity 
selection effect is indeed visible where the bottom panel under a wide range of $L_{\mathrm{bol}}$ displays flatter $f_{2.3}$ 
with $M_{\mathrm{BH}}$, or more chance of dust poor quasars at higher $M_{\mathrm{BH}}$ than from the upper panel under 
a fixed range of $L_{\mathrm{bol}}$. Therefore we expect the disconnection between $f_{3.5}$ or CF$_{hd}$ versus 
$M_{\mathrm{BH}}$ in the $z$\,$<$\,6 population of J10 or M11 to be, at least partially caused by the relatively broad 
range of $L_{\mathrm{bol}}$ that would smooth out the possible underlying correlation between $M_{\mathrm{BH}}$ and 
$f_{3.5}$/CF$_{hd}$. 

Likewise, the luminosity selection would act on the Eddington ratio as $f_{\mathrm{Edd}} \propto L_{\mathrm{bol}}/M_{\mathrm{BH}} 
\propto \sqrt{L_{\mathrm{bol}}}$, selecting more dust poor quasars at higher $f_{\mathrm{Edd}}$ than when volume limited. 
This time the positive $f_{\mathrm{Edd}}$--$p_{hdp}$ relation of Figure 12b is consistent with the $f_{\mathrm{Edd}}$--CF$_{hd}$ 
anti-correlation in Figure 12d, which can be explained by the stronger $f_{\mathrm{Edd}}$--CF$_{hd}$ 
anti-correlation than that of $L_{\mathrm{bol}}$--CF$_{hd}$ (e.g,. Figures 12d versus 11c). Furthermore, we expect the luminosity 
selection effect to be somewhat mixed by the redshift selection effect, due to the $z$--$p_{hdp}$ trend and the luminosity 
distribution of our sample such that higher luminosity sources are at higher redshift on average. We tested the redshift selection 
effect by drawing Figures similar to 12a and 12b at a fixed $L_{\mathrm{bol}}$\,=\,10$^{47-47.5}\,$erg\,s$^{-1}$ interval, 
while comparing the $p_{hdp}$ between 1\,$<$\,$z$\,$<$\,2 and 2\,$<$\,$z$\,$<$\,3. We find a higher normalization of $p_{hdp}$ 
at higher redshift, but the difference in the $p_{hdp}$ is smaller than when the redshift interval is fixed as one the two above, 
and instead adjacent subsamples in $L_{\mathrm{bol}}$ to 10$^{47-47.5}\,$erg\,s$^{-1}$ are compared. Therefore, the 
luminosity selection effect on $p_{hdp}$ in this work, is only mildly mixed with that from redshift selection.

As a way to remove the luminosity selection effect, M11 defined their hot dust poor quasars as the lower outlying 
objects in CF$_{hd}$, dependent on $L_{\mathrm{bol}}$. Therefore, the $p_{hdp}$ based parameter trends of dust poor quasars 
in M11 can be directly compared to the volume limited $p_{hdp}$ based trends in this work, as both studies are corrected 
for the luminosity selection, while the sample size of M11 is as large as ours such that they do not suffer from small 
number statistics. The $L_{\mathrm{bol}}$--$p_{hdp}$ trend would be different for the two studies though, as the definition 
of the dust-poorness in M11 inherently leads to a flat $p_{hdp}$--$L_{\mathrm{bol}}$ relation, while our result of the 
increasing $p_{hdp}$ with $L_{\mathrm{bol}}$ is free from such a selection. Apart from this exception, we expect the $p_{hdp}$ 
related parameter trends from M11 and this work to be consistent with each other, but we still witness somewhat conflicting 
results in $z$, $M_{\mathrm{BH}}$, and $f_{\mathrm{Edd}}$ spaces. However, the redshift range of 0.75\,$<$\,$z$\,$<$\,2 
covered by M11 is not as wide as that of this work, which according to Figure 11b, implies how they would miss up to an 
additional order of the increase in $p_{hdp}$ above $z$\,$=$\,2. Next, M11 did not find the $p_{hdp}$ to be depending 
on $M_{\mathrm{BH}}$ or $f_{\mathrm{Edd}}$ while we find negative and positive dependences on $M_{\mathrm{BH}}$ and 
$f_{\mathrm{Edd}}$, respectively. A caveat is that the dust poor selection of M11 passes through $\gtrsim\,$16\% of the 
sample, which is much higher than 0.6\% from this work, or $\sim$\,1\% from Figure 11b within the redshift interval 
0.75\,$<$\,$z$\,$<$\,2 matched with M11. From this, we suggest that the method of M11 to select low CF$_{hd}$ sources, 
may be choosing too many objects to be dust poor, making it hard to find distinctive properties of dust poor quasars from 
the rest of the sample. To summarize, we conclude that the differences in $p_{hdp}$ trends between M11 and this work, 
originate from the different redshift coverage and the selection criteria of dust poor quasars.

We note that the observational characteristics of dust poor quasars could be different in lower AGN luminosities, as our K--S 
probabilities in Table 3 are weaker at $L_{\mathrm{bol}}$\,$<$\,10$^{46.5}\,$erg\,s$^{-1}$. Combining our results with the 
study of Mor \& Netzer (2012, hereafter M12) helps to constrain the properties of fainter dust poor AGNs, since their sample are 
Seyfert 1 galaxies mostly covering $L_{\mathrm{bol}}$\,=\,10$^{44-46}$\,erg\,s$^{-1}$. Although their dust poor/total sample 
sizes are only $\sim$\,10 and 115, they do find low CF$_{hd}$ sources for narrow-line Seyfert 1s in low $M_{\mathrm{BH}}$ 
and high $f_{\mathrm{Edd}}$, consistent with our results at $>$\,10$^{46.5}\,$erg\,s$^{-1}$. Since the optical luminosities 
of AGNs in M12 are likely to be affected by host galaxy contamination (e.g., S11), it would be interesting for future studies 
to check whether the parameter trends of intermediate luminosity ($L_{\mathrm{bol}}\lesssim10^{46}\,$erg\,s$^{-1}$) dust 
poor AGNs as in M12, are valid after correcting for the host contamination in the optical luminosity. 

Physically connecting our observational results of luminous dust poor quasars, the $L_{\mathrm{bol}}$--$p_{hdp}$ relation 
is well explained by the receding torus model \citep{Law91}, such that more luminous quasars have smaller dust covering 
factors due to the torus being located far from the central light source. In addition, the link between FWHM--$p_{hdp}$ in 
Figure 13a is closely connected with the $M_{\mathrm{BH}}$--$p_{hdp}$ and $f_{\mathrm{Edd}}$--$p_{hdp}$ relations in 
Figures 12a and 12b, since quasars with narrow FWHM would be lower in $M_{\mathrm{BH}}$ and higher in $f_{\mathrm{Edd}}$ 
at a given $L_{\mathrm{bol}}$. This implies that quasars not only luminous but violent in growth and small in cumulative mass, 
or actively growing quasars, would have a higher chance of being dust poor. Since our dust poor quasars, 
although skewed to higher--$z$, are fairly spread within the $z$--$p_{hdp}$ relation, we refer to the dust poor epoch of 
quasars as positioned in a buildup state within individual black hole growth histories, rather than a specific global 
cosmic epoch (J10). 

In addition to the key trends, the $v_{offset}$--$p_{hdp}$ relationship for dust poor quasars preferring blueshifted broad 
emission, could be related to radiative outflows affecting the broad line emission (e.g., \citealt{Ric11}), which are necessary 
for radiative quasar feedback or QSO dust production through the expansion of broad line clouds (\S5.2). Alternatively, 
the blueshifts could be explained as recoiling black holes separated from the dusty torus \citep{Gue11}, but still the 
predicted probabilities of detecting kinematically offset AGNs are lower by several orders than the fraction of our dust 
poor quasars with large blueshifts. Meanwhile, the high FeII$/$MgII of dust poor quasars could be signs of metal seeds 
where the QSO dust can grow (\S5.2, \citealt{Pip11}), although it is difficult to make a clear interpretation from observed 
metallicities alone.

\subsection{The origin of dust poor quasars}
We now consider possible scenarios for the origin of dust poor quasars. First, we would like to list the explanations 
that the occurrence of dust poor objects are dependent on the geometry of the surrounding dusty structure. One of the 
descriptions is the receding torus model \citep{Law91} which predicts smaller dust covering factors at higher luminosities, 
for the obscuring structure to be located more outward. Our study provides further constraints to the model, as our dust 
poor quasars are not only more luminous, but also less massive and high in Eddington ratio. To fit our observational 
results the viewing angle of dust poor quasars could be considered as being observed from a face-on direction, because the 
unified model predicts obscuration from the dusty torus when observed through an inclined angle. Assuming that the broad 
line region follows the inclination of the torus \citep{Gas07}, and a simplified planar geometry for these AGN substructures, 
the observed line widths would meet FWHM$_{obs}$\,=\,FWHM\,$\sin i$, where $i$ is the inclination angle. Therefore, we 
may expect the broad FWHMs of dust poor quasars to be systematically narrower merely due to the projected $\sin i$ factor 
for a planar geometry. 

While this orientation based approach is capable of explaining many parameter space features (FWHM, $M_{\mathrm{BH}}, 
f_{\mathrm{Edd}}$) of dust poor quasars, the problem is that it cannot describe the redshift dependence of $p_{hdp}$, unless 
higher redshift quasars are systematically biased to lower inclination angle objects at given luminosity. Moreover, when the 
dusty torus is observed face-on, the unified model suggests more chances of being directly showered by radio jets (e.g., 
Figure 3 in \citealt{Ant93}), whereas in Table 3 we do not find dust poor quasars to be radio louder, although number 
statistics may not be secure. Several other geometric explanations to the origin of dust poor quasars include a low level 
misalignment of the torus with respect to the accretion disk \citep{Kaw11}, or mildly misaligned disks even without the 
torus structure \citep{Law10}. At this time where our observations are not able to strictly validate each geometric scenario, 
we leave it as an open question to explain dust poor quasars in a geometric sense, though we do stress the need for future 
models to be able to further explain our observational features (\S5.1). 

Second, in terms of physical origin of dust poor AGNs, that is, for intrinsic lack of the dust emission under the 
blowout process of dust, we would like to suggest in which stage dust poor quasars would lie if they follow existing 
AGN evolutionary scenarios. In regard to the triggering of quasar activity, we follow \citet{Tre12} to apply their 
results that within the luminosity limit of our study ($L_{\mathrm{bol}}$\,$>$\,$10^{45.7}\,$erg\,s$^{-1}$) more than 
50\% of quasars are triggered by major mergers. Merger triggered quasar activity are often observationally interpreted to 
initiate from the stage of luminous infrared galaxies, showing strong signatures of dense gas and dust up to kpc scales 
(e.g., \citealt{San88}; \citealt{Sur98}). Therefore, for the majority of luminous dust poor quasars to be originated from 
galaxy merging, not only the host galaxies are expected to have been dusty during the merging stage, but also the obscured 
host galaxies require feedback mechanisms that blow out the surrounding material to become more transparent to the 
quasar radiation alike normal quasar systems. This step is predicted in the merger driven AGN model of \citet{Hop08} as 
the blowout stage, where the gas and dust surrounding the quasar host galaxy are expelled.  

Depending on the extent of merger driven quasar feedback, red quasars are thought to be in the blowout stage 
where the surrounding dust have not yet been removed (e.g., Figure 5e in \citealt{Hop08}; \citealt{Urr08}), while we may 
now place dust poor quasars at the end of the blowout stage for the dust to have sufficiently been dispersed. Our study 
further helps to observationally constrain the evolutionary boundary of dust poor quasars as we find it to agree with the 
blowout phase predictions of rapidly growing BHs (low $M_{\mathrm{BH}}$, high $f_{\mathrm{Edd}}$) and intense feedback 
(high $L_{\mathrm{bol}}$ and blue UV continuum) from \citet{Hop08}, and with observations of high $f_{\mathrm{Edd}}$ objects 
within the red quasar sample of \citet{Urr12}. At the same time however, dust poor quasars are closer to normal quasar 
phases as their accretion disk dominated SEDs indicate the nucleus to be less obscured from its surrounding, compared to 
red quasars. Therefore, we consider it the most plausible to assign dust poor quasars in between the blowout and traditional 
quasar phases, where dust poor quasars can be explained to have just become unobscured as they went through intense 
feedback during the rapid black hole growth, but are still relatively low in $M_{\mathrm{BH}}$ and high in $f_{\mathrm{Edd}}$ 
on their way of becoming normal quasars. When our dust poor fraction of 0.6\,\% is translated into the visible dust poor 
timescale, it becomes $\sim$\,0.1--1\,Myr from the visible timescale of quasar activity (\citealt{Mar04}; \citealt{Hop05}). 
Therefore, the short visible timescale possibly implies either that the dust can form efficiently after the short/intense 
dust dispersing part of the blowout phase, or that it is difficult for typical quasars to clear off the surrounding dust 
to a sufficient extent.

The approach to clarify the evolutionary state of dust poor quasars from the key parameter trends, would be strengthened 
when the redshift evolution of $p_{hdp}$ from this work is further explained. At high redshift the amount of dust observed 
in quasars or GRBs can be explained by dust producing sources mainly from supernovae and the most massive AGB 
stars, whereas QSO dust\footnote{We use the term QSO dust for the dust production to originate from the quasar activity, 
and to distinguish from stellar related sources.} or contributions from less massive AGB stars are relatively limited (e.g., 
\citealt{Pip11}, \citealt{Jan11}). The restricted dust production routes within the short age of high redshift quasar systems 
therefore, could be the cause of the $z$--$p_{hdp}$ relation of dust poor quasars as their progenitors may not have 
enriched enough dust to survive under the presence of AGN feedback, observed to be in action up to the early universe 
(e.g., \citealt{Mai12}). Hence, the increased fraction of dust poor quasars at higher redshift is consistent with the current dust 
formation model predictions, and with observations of bluer UV continuum slopes of inactive galaxies at high redshift (e.g., 
\citealt{Bow09}). 

Stemming from the evolutionary model for dust poor quasars, the fact that we find BAL quasars in our dust poor sample has 
interesting implications about the dust origin of BAL quasars. Utilizing the BAL flag in S11 based on the criterion from 
\citet{Gib09}, we find the BAL fraction of 4.3\,$\pm$\,1.5\% from the dust poor sample, to be similar or only slightly 
higher than 3.2\,$\pm$\,0.1\% from the entire sample. As stated earlier, dust poor quasars have a strong blue UV--optical 
continuum. If the BAL is caused by an orientation towards absorption (e.g., \citealt{Elv00}), BAL quasars would not be 
included in the dust poor sample since the dusty structure would obscure the UV--optical light too. The evolutionary 
model for dust poor quasars seems to provide a natural explanation for the dust in BAL, supporting an evolutionary model 
of the BAL phase to lie in between luminous infrared galaxies and normal quasars (e.g., \citealt{Bri84}; \citealt{Lip06}).

\subsection{The future of dust poor quasars}
Having considered dust poor quasars to possibly be explained by geometric or evolutionary models, to be observed by an 
unobscured orientation or to have undergone a duration of strong feedback, we would finally like to comment on their close 
future by comparing with average optically selected quasars. If dust poor quasars are originated from a geometric reason, 
we may expect the covering factor to become larger at later quasar phases when they become quieter in luminosity or 
Eddington ratio, as the covering factors are anti-correlated with $L_{\mathrm{bol}}$ or $f_{\mathrm{Edd}}$ (\citealt{Law91}; \citealt{Kaw11}).

On the other hand, assuming an evolutionary origin of dust poor quasars, we need to consider dust formation mechanisms 
during the AGN evolution. Although our observational explanation of dust poor quasars to be rapidly growing (\S5.1) support 
them to be younger than ordinary quasars, the evolutionary models predict the other way round, for quasars to become more 
dust poor along the duration of its activity. This is due to the solely destructive nature of AGN feedback, to blow out the 
surrounding dust (e.g., \citealt{Haa03}; \citealt{Hop08}). Thus, assuming that the evolutionary paths for dust poor and 
ordinary quasars are aligned, in other words, considering the dust poor phase to be general within the lifetime of optically 
selected quasars, the only way to reconcile with the observations is to add to the models constructive feedback from the 
AGN activity itself, that is to say, dust formation mechanism at the center of quasars. This idea makes sense that it meets 
the time causality to enrich the dusty tori as the system accumulates its black hole mass. Our rapid growth scenario 
therefore naturally supports the model production of QSO dust, possibly during the free expansion of broad line clouds 
\citep{Elv02}, which is in fact likely to be the dominant source of dust formation in the inner galaxy once AGN feedback 
turns effective \citep{Pip11}. To summarize, revising the evolutionary model for luminous quasars passing the dust poor 
phase, dust blowout in the inner kpc scale of the galaxy, is followed by dust formation from the central pc scale of the AGN. 

Still, in the cases of powerful AGN feedback, one could imagine the intense radiation to not only remove the galaxy scale 
dust, but destroy the freshly formed QSO dust at the center of active galaxies such that dust poor quasars remain 
dust poor. This may not be the case for $L_{\mathrm{bol}}<2\times10^{47}\,$erg\,s$^{-1}$ quasars though, since the 
radiative flux density shining the QSO dust forming region is weaker than that around giant stars \citep{Elv02}. Considering 
that 75\% of our dust poor quasars are less luminous than this limit, the extreme case of continued dust destruction can 
mostly be rejected as the fate of dust poor quasars. Therefore, provided that the dust production is effectively working, 
we suggest the near future of dust poor quasars to be richer in dust including the contributions from QSO dust, which is 
against observation based scenarios where ordinary quasars become dust poor objects (\citealt{Haa03}; \citealt{Hao12} 
though for a fainter luminosity range), but supports dust enrichment within the torus (\citealt{Elv02}; J10) or circumnuclear 
regions (e.g., \citealt{Sim07}) of active galaxies during its activity. 

But what if dust poor quasars are not causally related to ordinary quasars? It could be the case where dust poor quasars 
are indeed rare objects out of AGN evolution, to be placed in intrinsically dust poor environment within their host galaxies 
and/or the QSO dust formation is weak, such that they do not become as dust rich as ordinary quasars during the quasar 
phase. This idea may be supported by observations of early-type AGN host galaxies, where the ratio of dust mass over 
4.5\,$\mu$m luminosity representing the stellar mass (e.g., \citealt{Jun08}), is widely ranged at given 4.5\,$\mu$m 
luminosity \citep{Mar13}. Within this framework, the relative lifetimes of each AGN evolutionary stage based on obscuration, 
could be scattered due to the variety of dust content within each host environment. For example, although the AGN 
evolutionary models predict obscured AGNs to be younger than the unobscured, under an especially low dust-to-stellar 
mass ratio environment of the host, AGNs would quickly shine optically luminous without a typically long duration being 
obscured (e.g., \citealt{Hop05}). Thus, even if dust poor quasars may be explained to lie in a rare dust poor environment, 
it shifts the visible dust poor lifetime to the relatively earlier growing state of the black hole, still consistent with 
our rapid growth explanation (\S5.1).

\section{Conclusion}
Out of a large area, multi-wavelength sample of luminous quasars, we identified 233 (0.6\%) objects that satisfy the hot 
dust poor criterion analogous to that of \citet{Jia10}. The selected dust poor quasars are weak in both NIR/MIR-to-optical 
flux ratios, and show a blue continuum with the mean slope of $\alpha$\,$\sim$\,0.1 from UV through NIR. We calculated 
the fraction of dust poor quasars in four bolometric luminosity bins, and find statistical preferences of dust poor quasars 
to be more abundant in higher redshift, and to have lower black hole mass, higher Eddington ratio, blueshifted and narrower 
broad line, and higher FeII/MgII ratio, at a given luminosity. The results show that dust poor quasars are rapidly growing 
population in terms of $M_{\mathrm{BH}}, f_{\mathrm{Edd}}$, while the rapid growth of black holes and the dust poorness 
should be linked closely with the evolution of quasars as a function of redshift.

To explain the observational characteristics of dust poor quasars, we suggest a scenario in which dust poor quasars are 
transient phenomena during an evolutionary process, when they became unobscured by merger driven AGN feedback. 
Luminous quasars are often triggered by violent major merging \citep{Tre12}, involving large extinctions from 
their host galaxies (e.g., \citealt{San88}). From the onset of strong AGN activity, radiative feedback blows out \citep{Hop08} 
the surrounding galactic dust to become dust poor quasars, for a very short period if they produce QSO dust (\citealt{Elv02}; 
\citealt{Pip11}), or for longer if they are in an especially dust poor environment. The redshift 
evolution of the dust poor fraction is an indication of the dust content at different redshift, where the dust poor phase 
is more easily identified and lasts longer in higher redshift, since the elements that make up dust grains are scarce 
early in the universe. An alternative scenario would be to explain dust poor quasars to be a distinct population with 
small covering factors, due to the geometry/orientation of the obscuring material. Such models (\citealt{Law91}; \citealt{Law10}; 
\citealt{Kaw11}) are consistent with most of the observed properties, but not with the redshift evolution of the dust 
poor fraction. 

We find the short timescale from the rare population of dust poor quasars, to provide meaningful evolutionary predictions. 
Deep and resolved imaging of the AGN host will tell us if these dust poor quasars are closely linked to the evolutionary 
phase, as the merging features would be clearer at earlier times than the rapid fading of tidal features \citep{Hop08}. 
Besides, extrapolating the rough relation $p_{hdp} \propto (1+z)^{2.34}$ (\S4.2) up to the early universe, we expect 
$p_{hdp}$\,=\,18\% and 37\% for luminous quasars at $z$\,$=$\,7 and 10. Therefore, discovery of the highest redshift 
quasars will help us know whether the higher incidence of dust poor quasars, blue in UV continuum, boost the quasar 
contribution in reionizing the universe (e.g., \citealt{Fan06}), or if they are born in extremely dust poor galaxies as hinted 
by the UV slopes of low luminosity galaxies at $z$\,$\sim$\,7 (e.g., \citealt{Bow10}). Future deep multi-wavelength studies 
of dust poor AGNs should place better constraints on the multi-temperature dust emission of the AGN-host system and the 
redshift evolution of these objects.

\acknowledgments

This work was supported by the National Research Foundation of Korea (NRF) grant, No. 2008-0060544, funded by the Korea 
government (MSIP). We thank the referee's helpful comments to improve the paper, and the CEOU team members for their contribution in taking the UKIRT data.
%GALEX
Based on observations made with the NASA Galaxy Evolution Explorer. 
GALEX is operated for NASA by the California Institute of Technology under NASA contract NAS5-98034.
%2MASS
This publication makes use of data products from the Two Micron All Sky Survey, 
which is a joint project of the University of Massachusetts and the Infrared Processing 
and Analysis Center/California Institute of Technology, funded by the National Aeronautics and 
Space Administration and the National Science Foundation.
%UKIDSS
This publication makes use of data products from the United Kingdom Infrared Deep Sky Survey, and the data taken with the 
UKIRT. The United Kingdom Infrared Telescope is operated by the Joint Astronomy Centre on behalf of the Science and 
Technology Facilities Council of the U.K. The UKIDSS photometric \citep{Hew06} data are from WFCAM imaging \citep{Cas07}, 
where the data are processed (Irwin et al. 2009, in prep), calibrated \citep{Hod09}, and science archived \citep{Ham08}.
%WISE
This publication makes use of data products from the Wide-field Infrared Survey Explorer, 
which is a joint project of the University of California, Los Angeles, and the Jet Propulsion 
Laboratory/California Institute of Technology, funded by the National Aeronautics and Space Administration.

\end{document}